\documentclass[10pt,journal,compsoc]{IEEEtran}
\ifCLASSOPTIONcompsoc
  \usepackage[nocompress]{cite}
\else
  \usepackage{cite}
\fi
\IEEEoverridecommandlockouts
\usepackage{cite}
\usepackage{amsmath,amssymb,amsfonts}
\usepackage{algorithmic}
\usepackage{graphicx}
\usepackage{textcomp}
\usepackage[table]{xcolor}
\usepackage{graphicx}
\usepackage{lscape}
\usepackage{multirow}
\usepackage{rotating}
\usepackage{subfigure}
\usepackage{afterpage}
\usepackage{tcolorbox}
\usepackage{url}
\usepackage{float}
\usepackage{makecell}
\usepackage{verbatim}

\newcommand{\todo}[1]{}
\renewcommand{\todo}[1]{{\color{red} TODO: {#1}}}

\def\BibTeX{{\rm B\kern-.05em{\sc i\kern-.025em b}\kern-.08em
    T\kern-.1667em\lower.7ex\hbox{E}\kern-.125emX}}
\begin{document}

\title{An Empirical Study on How Well Do COVID-19 Information Dashboards Service Users' Information Needs}

%\author{Xinyan Li, Han Wang, Chunyang Chen% <-this % stops a space
%\IEEEcompsocitemizethanks{\IEEEcompsocthanksitem Xinyan Li, Han Wang, Chunyang Chen
%are with Faculty of Information Technology, Monash University, Australia. E-mail: xlii0171@student.monash.edu, han.wang1@monash.edu, chunyang.chen@monash.edu.}}
% note need leading \protect in front of \\ to get a newline within \thanks as
% \\ is fragile and will error, could use \hfil\break instead.

\author{Xinyan~Li,
        Han~Wang,
        Chunyang~Chen,
        and~John~Grundy
\IEEEcompsocitemizethanks{\IEEEcompsocthanksitem Xinyan Li, Han Wang, Chunyang Chen, John Grundy
are with Faculty of Information Technology, Monash University, Australia. Xinyan Li and Han Wang are co-first authors of this paper, Chunyang Chen is the corresponding author.\protect\\
E-mail: xlii0171@student.monash.edu, \{han.wang, chunyang.chen, john.grundy\}@monash.edu. }}

\IEEEtitleabstractindextext{%
\begin{abstract}
%\chen{To be updated!}
The ongoing COVID-19 pandemic highlights the importance of dashboards for providing critical real-time information.
In order to enable people to obtain information in time and to understand complex statistical data, many developers have designed and implemented public-oriented COVID-19 ``information dashboards" during the pandemic.
However, development often takes a long time and developers are not clear about many people's information needs, resulting in gaps between information needs and supplies. According to our empirical study and observations with popular developed COVID-19 dashboards, this seriously impedes information acquirement.
Our study compares people's needs on Twitter with existing information suppliers. We determine that despite the COVID-19 information that is currently on existing dashboards, people are also interested in the relationship between COVID-19 and other viruses, the origin of COVID-19, vaccine development, fake new about COVID-19, impact on women, impact on school/university, and impact on business. Most of these have not yet been well addressed. 
We also summarise the visualization and interaction patterns commonly applied in dashboards, finding key patterns between data and visualization as well as visualization and interaction. Our findings can help developers to better optimize their dashboard to meet people's needs and make improvements to future crisis management dashboard development.

\end{abstract}

}

\maketitle

\IEEEdisplaynontitleabstractindextext
\IEEEpeerreviewmaketitle

\IEEEraisesectionheading{\section{Introduction}
\label{sec:introduction}}

\begin{table*}[h]
\centering
\resizebox{\textwidth}{!}{%
\begin{tabular}{llll}
\hline\hline
Dashboard name & URL                                           & Codename                                    & Scope \\
\hline\hline
COVID-19 in Australia Real-Time Report &
  https://covid-19-au.com/ &
  COVIDAu & National\\
Global COVID-19 Tracker \& Interactive Charts  & https://coronavirus.1point3acres.com/                                      & 1Point3Acres      &   National      \\
COVID-19 Dashboard by \\
Johns Hopkins University (JHU)  & https://coronavirus.jhu.edu/                                       & JHU &    Global       \\
DXY, DX Doctor COVID-19\\ 
Global Pandemic Real-time Report  & https://ncov.dxy.cn/ncovh5/view/pneumonia                           & DXY          &     National     \\
WHO Coronavirus Disease (COVID-19) Dashboard  & https://covid19.who.int/  & WHO            &   Global       \\\hline\hline
\end{tabular}%
}
\caption{Dashboards and names used in text}
\vspace{-8mm}
\label{tab:dashboards&codename}
\end{table*}

\IEEEPARstart{C}{oronavirus} disease 2019 (COVID-19)~\cite{zhou2020pneumonia} is an infectious disease caused by severe acute respiratory syndrome coronavirus 2 (SARS-CoV-2).
Since its first being reported in December 2019, COVID-19 has spread quickly around the world, and becomes a global pandemic. There were more than 195 million confirmed cases and over 4 million lives lost by far. 
%188 countries and regions have been effected and the United States have the largest confirmed case numbers followed by India and Brazil. 
Governments from different countries and states have taken actions to try to avoid the transmissions such as lockdown, border closure, claim social distance. And human lives have been severely impacted due to the pandemic. 
People are in desperate need of the latest information such as policies, and the states in their local area from many different sources, i.e. social media, informative dashboards, government releases, TV, newspaper. 

However, current information sources have a number of problems when providing COVID-19 information web services. First, the information from government or traditional media (i.e., TV and newspaper) is not frequently updated. Second, different layers of the government (state and federal government) may provide contradictory information. Also, there are also many rumors spread on social media~\cite{brindha2020social}, which makes it difficult for people to know who and what to trust. Finally, the current information about COVID-19 is fragmented. It takes effort for people to aggregate the information they need to see from different places.

A dashboard is a visual display of the most important information needed to achieve one or more objectives~\cite{few2007dashboard}. Information is typically consolidated and arranged on a single screen to be monitored at a glance.
COVID-19 dashboards have emerged for collecting COVID-19 information in a one-stop sharing site and providing services to address the problems we discussed above. They offer real-time statistics about different aspects of the COVID-19 virus, use different visualizations to help understand the trends of the pandemic, compare situations in different regions, and acknowledge users areas that are being severely infected. 

Since these dashboards are hosted online, developers are able to update them multiple times per day to serve the latest data to public once they are available. They have good mobility as people can access it anytime at anywhere. Some of them allow users to subscribe for the latest updates. Unlike social media forums where everyone can post, dashboards are only maintained by the developers, so the chance of encountering confused data or rumors is reduced.

\begin{figure}
\centering
\subfigure
[JHU dashboard] 
{\includegraphics[width=4cm]{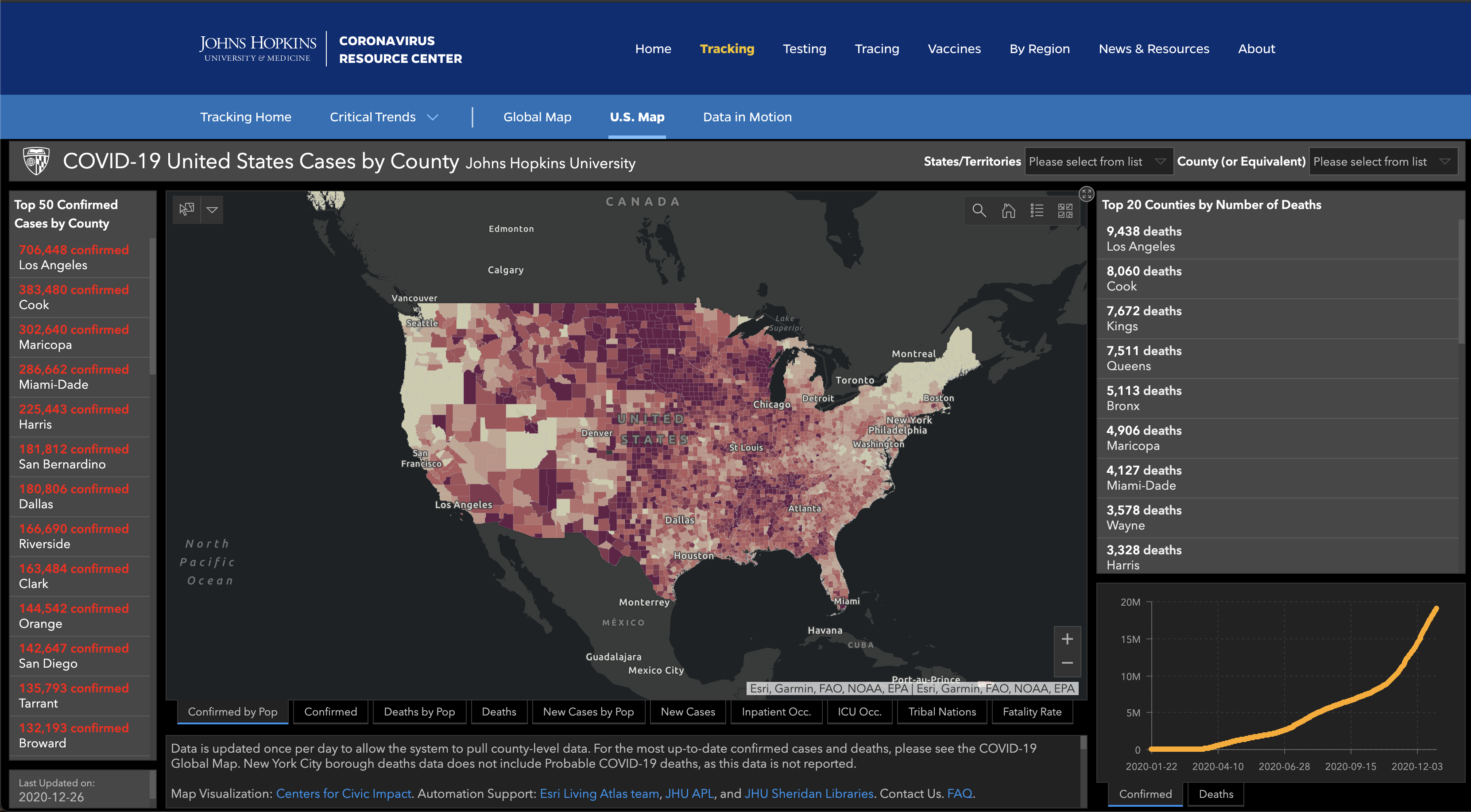}} 
\subfigure[WHO dashboard]{\includegraphics[width=4cm]{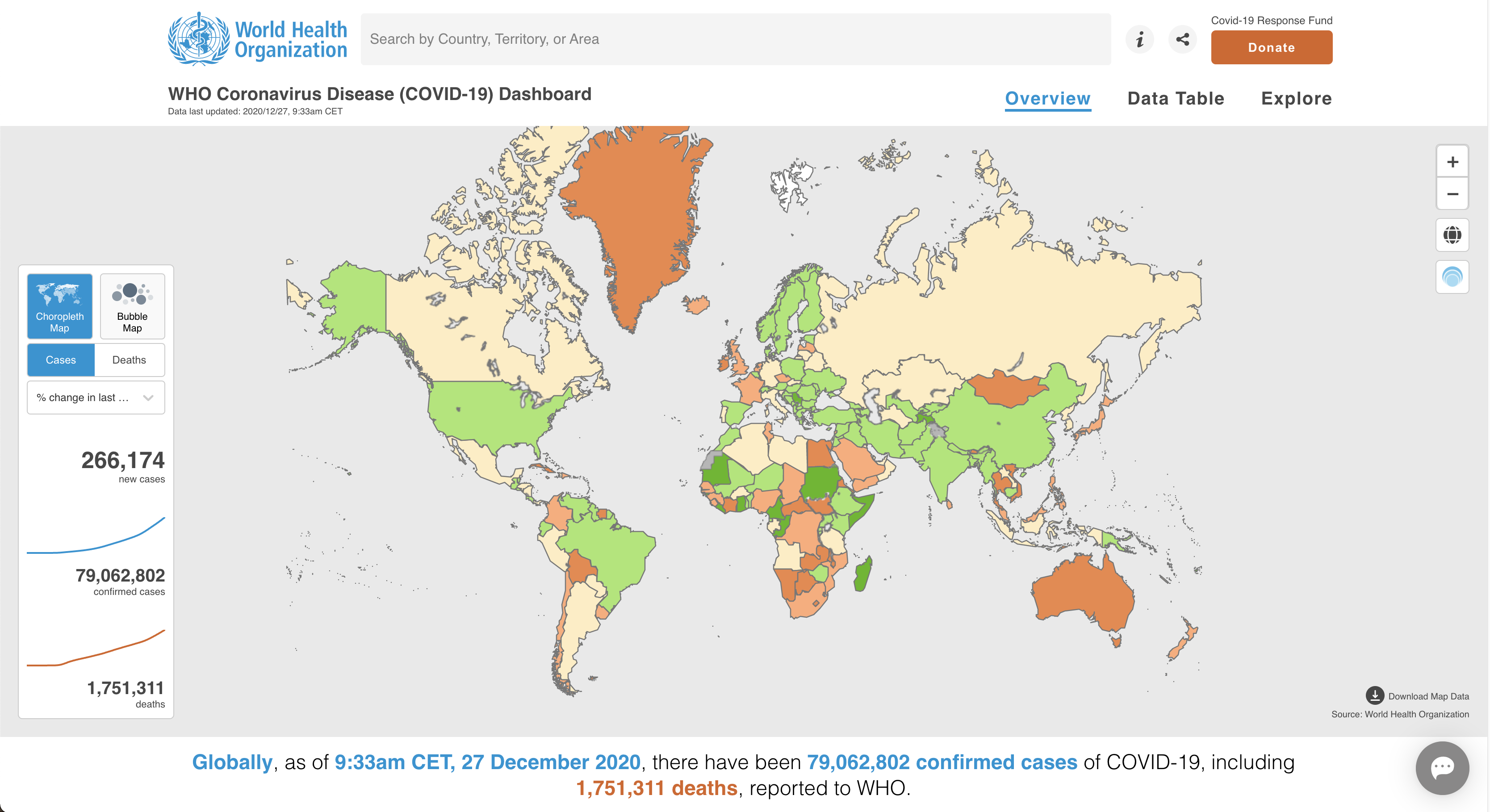}}
\subfigure[DXY dashboard]{\includegraphics[width=2.3cm, height=3.4cm]{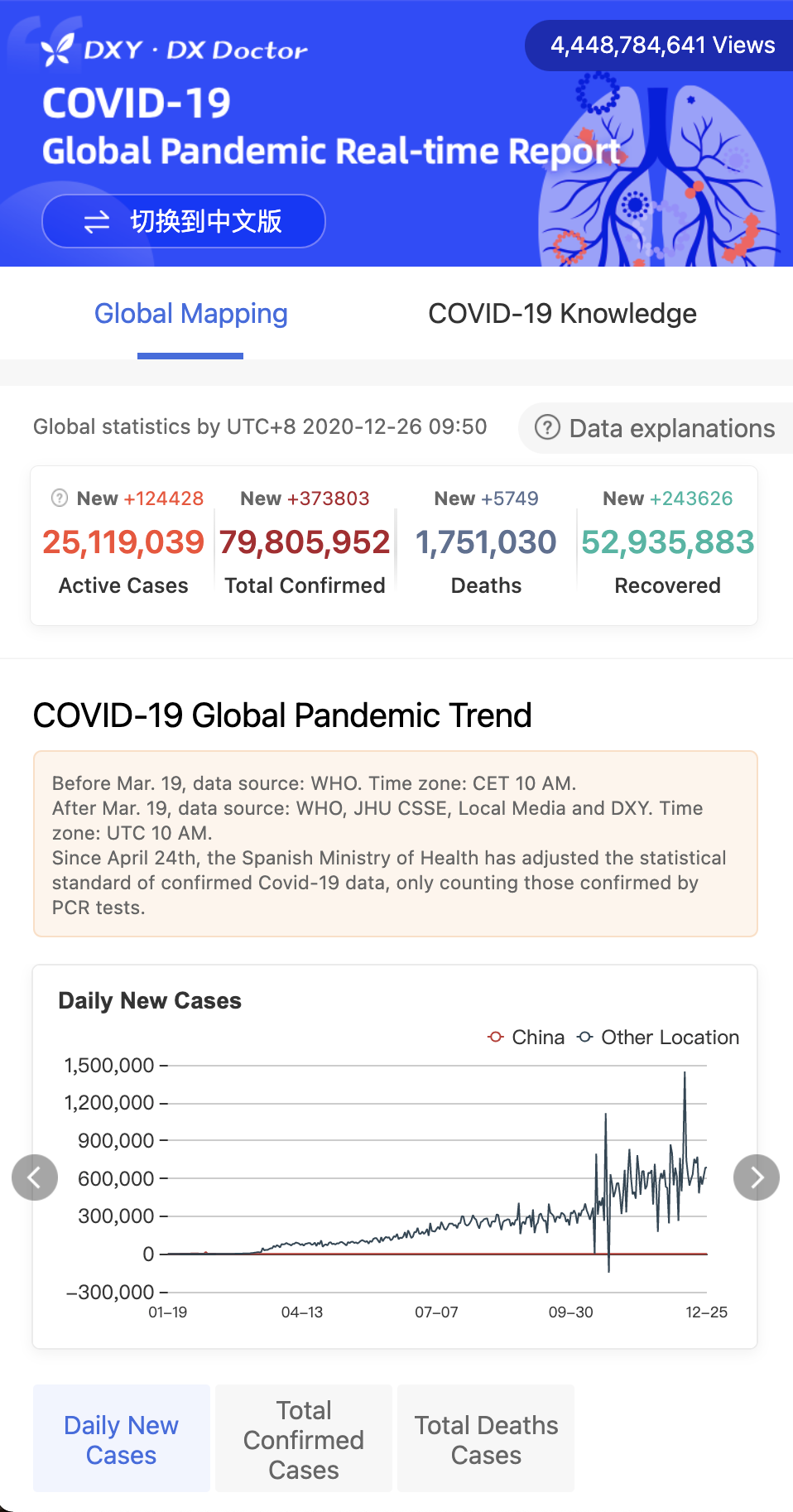}}
\subfigure[1P3A dashboard]{\includegraphics[width=2.3cm, height=3.4cm]{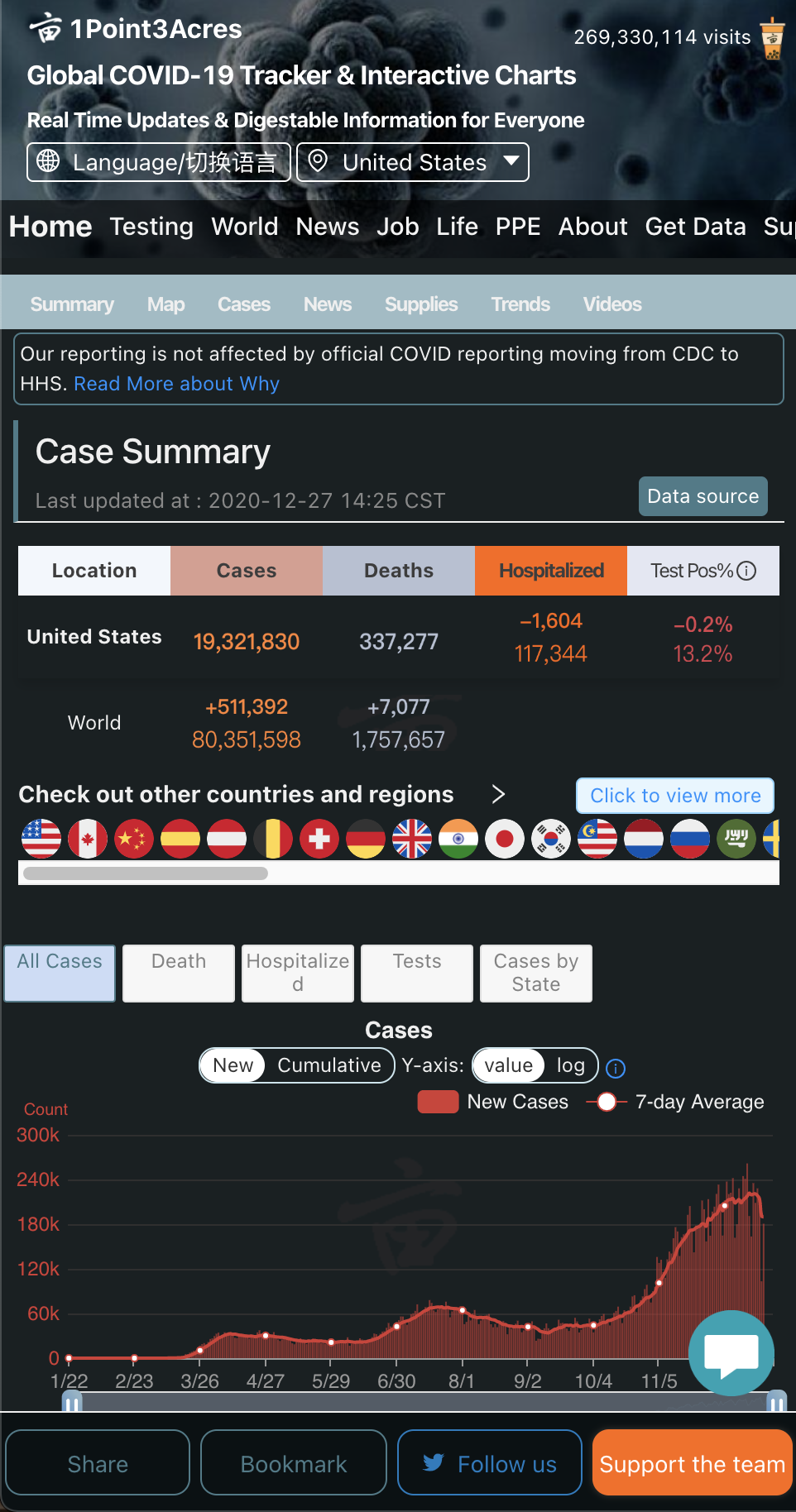}}
\subfigure[COVIDAu dashboard]{\includegraphics[width=2.3cm, height=3.4cm]{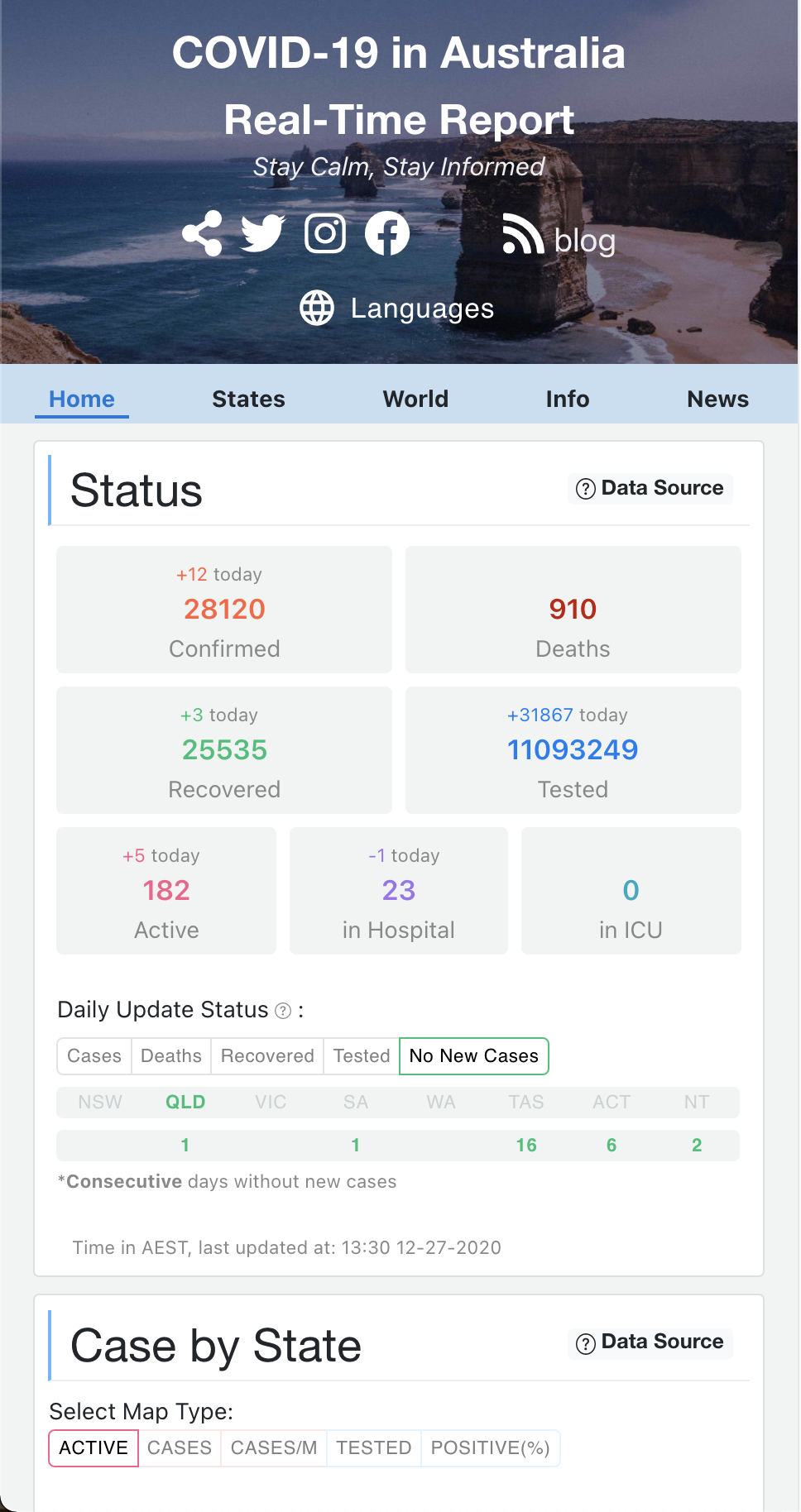}}
\vspace{-3mm}
\caption{Screenshots of different popular COVID-19 dashboards} 
\vspace{-6mm}
\label{fig:1} 
\end{figure}

There are some works related to dashboard development~\cite{few2007dashboard, pearcy2018computer}. However, no previous works provide comprehensive instructions on building dashboards for crisis events. Aside from basic stats (i.e., confirmed case, death), different dashboard websites have various types of data. Moreover, even for the same type of data, there are different 
approaches to visualize them. Some of the visualizations are useful whereas some are irresponsible use of visualization that may cause public fear, for example using sensationalist colours in the maps~\cite{arcgisBlog:online}.
It is unclear what types of data that dashboard users want to learn about the most, and what is a proper way to visualize the pandemic data so that the public can better understanding the situation.
To address this, we carried out an empirical study of COVID-19 information dashboards with the aim of forming a more systematic way of building such dashboards for crisis events to serve the public needs. We propose four research questions and summarize the questions and findings.
We analyze tweets related to the COVID-19 topics, compare features of various COVID-19 dashboards, and build a knowledge taxonomy. We further discuss the implications that we find during the research. We hope that this work can help the current COVID-19 information dashboard provide a better web service to the public and be helpful for the development of a similar crisis event dashboard in the future.

The key contributions of this work include:

\begin{itemize}
    \item This is the first empirical study on information dashboards that focuses on serving the information needs for the COVID-19 pandemic.

	\item By comparing the information requirements from social media and the contents from existing COVID-19 dashboards, we find gaps between the current information dashboards and the public's needs. We identify two critical themes of information that people frequently ask about on social media but do not appear on dashboards.
	
	\item By analysing real users’ feedback about COVID-19 dashboards on Twitter, we identify the pros and cons of different dashboards and analyze the reasons behind these.
	
	\item We discuss lessons learnt from different visualizations on dashboards that could be adopted for rapidly developing other emergency information dashboards in the future. 
	
\end{itemize}

%\todo{paper overview}

We first discuss the motivation of this research and specify its three key research questions (RQ) (Section 2). We then describe the method for data collection and analysis (Section 3). We illustrate the approach and results we have obtained for each RQ (Sections 4-7). The discussion of results is discussed (Section 8) followed by threats to validity (Section 9) and key related works (Section 10). We conclude the paper and discuss possible future works (Section 11).

\section{Motivation}

%\todo{

%Suggest move Fig 1 here, overview the key dashboards shown, and summarise their key problems; issues.  Perhaps then identify sthe key RQs and short number of requirements for improved dashboards

%before do Approach in next section

%}

A large number of COVID-19 information dashboards have been published to serve the public information needs. 
For instance, the Johns Hopkins University COVID-19 dashboard has become an essential resource for tracking the scope and scale of the pandemic, not just for media outlets but also for clinicians and government officials~\cite{web:JHUHelp}. 
DingXiangYuan (DXY), a Chinese organization, builds a dashboard to track the COVID-19 virus in China and has more than 4 billion visits across the world. 
We have summarised some of the more popular COVID-19 dashboard websites in Fig~\ref{fig:1}. 
Among them, the COVIDAu dashboard is developed by the authors of this paper. 
Our dashboard primarily focuses on the pandemic in Australia, and its purpose is to keep the Australians informed about the latest government-provided information. 
We also include the 1point3acres dashboard, which is built to serve the U.S. and Canada users, and the World Health Organization (WHO) COVID-19 dashboard, which is used to report global pandemic trends.

Despite the high volume of visits to these dashboards, we found some questions during the usage and development.
For instance, during our development of the COVIDAu dashboard, we received many requests for different types of data and visualizations, e.g.,``\textit{Also, can you also show the cases summary per suburb.}" and ``\textit{Are you able to show \# of tests per day conducted (preferably per state) in the historical section?}". 
This makes us wonder what information people want to know that hasn't yet been put into the dashboard.
When looking at COVID-19 dashboards, we find that basic data like total cases or death numbers are developed with different visualizations.
For instance, one dashboard uses both line charts and histograms on total cases data, whereas another dashboard applies map visualization on the same type of data. 
As there is no such pandemic like the COVID-19 that makes such enormous impacts globally, no related works have been found on how to improve information services on these pandemic dashboards.

\subsection{Research Questions}

%\todo{Move to Motivation section as suggested above?}

To present data that meets public information needs and improves the current COVID-19 dashboards, we explore their information needs and existing information supply by carrying out an empirical study of multiple COVID-19 dashboards.
Developers can use the findings to improve these dashboards further and prepare the community for the next potential pandemic.
%In this research, our goal is to provide an empirical study that evaluates the information needs to build a COVID-19 information dashboard. 
Below we present and discuss the key research questions this study attempts to answer.
We discuss why we raise the questions, our plan to answer them, and the benefits of answering each of the questions. 

\begin{center}
 \begin{tcolorbox}[arc=0pt,boxrule=1pt,colback=white!5!white,colframe=black!95!black,bottom=-0.5pt,top=-0.5pt]
\textbf{RQ1.} What are people's information needs about COVID-19?
\end{tcolorbox}
\end{center}

The use of dashboards has become popular and they have been applied in many situations, such as tracking disease spreading or the changing in business market~\cite{TheImpor41:online}. 
Given that the COVID-19 virus spreads rapidly and people want to know the latest status, we propose making a dashboard that provides real-time COVID-19 information would needs to collect the information that the masses want.
However, traditional requirements collection methods such as surveys are too cumbersome~\cite{dale1982alternative}.
After investigation, we notice that people are very enthusiastic about discussing the COVID-19 on Twitter~\cite{singh2020first}. 
% Due to its high volume of data, we can not manually read all the tweets and analyze what they need. 
We therefore, use automated data analysis to analyze a number of tweets to get people's views on their COVID-19 pandemic information needs. 
RQ1 aims to extract information that people desire to know based on collecting COVID-19 related inquires on Twitter. 
We use topic modeling to extract topics of information, as this avoids duplication of data.

\begin{center}
 \begin{tcolorbox}[arc=0pt,boxrule=1pt,colback=white!5!white,colframe=black!95!black,bottom=-0.5pt,top=-0.5pt]
\textbf{RQ2.} What are the information supplies from existing COVID-19 information dashboards?
\end{tcolorbox}
\end{center}

The dashboard is a common and popular information management tool that is often used to track and visualize major public events and communicate important information at a glance~\cite{few2006information}. 
Since the COVID-19 coronavirus was discovered, many dashboards have been developed, including the dashboard developed by the authors. 
Nevertheless, there has no research summary of the information in the existing dashboards. 
Hence, we propose our second research question to explore this field. 
Comparing the findings from RQ1 and RQ2 can reveal the gap between existing information supply and people's expressed requirements.
% The answer to RQ2 can also help developers pick up data for crisis dashboards in the future quickly.

\begin{center}
 \begin{tcolorbox}[arc=0pt,boxrule=1pt,colback=white!5!white,colframe=black!95!black,bottom=-0.5pt,top=-0.5pt]
\textbf{RQ3.} How should COVID-19 related information be better visualized?
\end{tcolorbox}
\end{center}

%\todo{I wonder about generalising this to "crises event information dashboards" in general?  Or just discuss in Discusion/Future work??}

Visualization and interaction are essential for information exploration in dashboards.
Previous research has discussed data type taxonomies for information visualization, and discovered that a successful commercial product has multiple data types~\cite{shneiderman1996eyes}. 
Still, there is no research focusing on the information visualization needs and approaches for COVID-19 data. 
Hence, we raise the RQ3 to analyze data visualizations. 
By summarising the visualization and interactions on the existing COVID-19 dashboards, RQ3 provides a guideline for further emergency information dashboard development, particularly for dashboard developers.

%\todo{Ideally it would be good to collect user feedback on the dashboards in use if possible... eg. 

%RQ4  What are users views on current COVID-19 Dashboards?
%}

\begin{center}
 \begin{tcolorbox}[arc=0pt,boxrule=1pt,colback=white!5!white,colframe=black!95!black,bottom=-0.5pt,top=-0.5pt]
\textbf{RQ4.} What are users' views on current COVID-19 Dashboards?
\end{tcolorbox}
\end{center}

The purpose of developing a COVID-19 dashboard is to help people understand the epidemic information quickly and efficiently. 
Therefore, users' comments can be very important for developers to improve the dashboard services in future development.
We proposed the RQ4 to study users' comments on COVID-19 dashboards.
By exploring users' comments on Twitter, we summarised the positive and negative comments.
Besides, we also discuss possible reasons that caused the negative comments.

\section{Method}

We adopt a mixed-methods approach by conducting Twitter information collection and existing COVID-19 dashboard analysis. 
The following section discusses how we retrieve data from Twitter and the methods we chose to attempt to answer each research question.

\subsection{Data Collection}
\label{sec:open-coding}
%\todo{Suggest explain data analysis @ end of this section too}

We collected tweets with COVID-19 related hashtags from 22${^{nd}}$ March to 10${^{th}}$ June 2020 to try to analyze users' expressed COVID-19 information needs. 
In total, we collected 56M tweets by applying an advanced search for specific COVID-19 hashtags. 
Some of the hashtags are from Breslin et al.~\cite{breslin2020we} and our research team summarized the others.
\label{sec:hashtags}
All of them are related to \textit{\#COVID-19}. 
In detail, we have collected 1.54M tweets with \textit{\#CoronaVirusUpdate}, 7.9M tweets with the hashtag \textit{\#Corona}, 14.3M tweets about \textit{\#Coronavirus}, 2.8M tweets have \textit{\#Covid} hashtag, and 20.7M tweets contains the hashtag \textit{\#Covid\_19}.
We stored the tweets in CSV files and created a script to filter out non-English tweets. 
Finally, we obtain 48M English tweets related to COVID-19 topics.

To comprehend dashboards that developers have produced, we select five dashboards, as can be seen in Table~\ref{tab:dashboards&codename}: the COVIDAu dashboard, a dashboard that has been built by the authors and focuses on COVID-19 in Australia; the 1Point3Acres dashboard, which serves the needs of people mainly from the US; the JHU dashboard, a dashboard that provides a live update for over 190 different countries and areas; the DXY dashboard, one of the earliest COVID-19 dashboards; and the WHO COVID-19 dashboard. 
These five dashboards are popular in their countries/regions and have millions of visitors in total.
The dashboards we selected not just have the ones that serve for specific countries/areas (China, North America, and Australia), but also include ones like JHU and WHO that focus on global data. 
% We selected the dashboard from China as it reports the first COVID-19 cases, dashboards from North America as they have the most confirmed cases in the world, and the COVIDAu dashboard as it was developed by the authors and we have first-hand experience and data from it.
% Our research covered the dashboards of countries and regions with more severe epidemics. 
% For example, in the country where reported the first COVID19 case, China, we chose the popular dashboard in the local area.

\subsection{Data Analysis}

%\todo{Need more details on coding e.g. why this approach chose, how many coded, cross-validation etc. Here or in answer to specific RQ??}

After determining the dashboards to analyze, we adopted an open-coding method~\cite{opencodi12:online} to collect and integrate information about them. 
Our research team divided into two groups.
Each group collected all the information displayed on those dashboards in the form of texts or charts to help users better understand data.
% All the information we collected was displayed on the dashboard in the form of text or charts to help users understand, for instance, the confirmed cases on the case map. 
The two groups marked the unique information that only appears in one dashboard to facilitate subsequent dashboard difference comparisons.
After that, the groups categorized the dashboard information into different themes and compared it with people's needs expressed via the Twitter analysis. 
We also collected visualizations and interactions from these dashboards.
Two teams manually checked the charts in each dashboard and the interaction methods supported by those charts. 
%During the research process, we held a meeting every week to check and organize the information to ensure that all the data we collected was accurate.
After the initial coding, the two teams met and discussed any discrepancies and new information categories until they reached consensus.
All the information we collected on each dashboard is stored in Google sheets\footnote{Google sheets URL: https://cutt.ly/Kjuzb5J}.

%\todo{Is there any empirical study of the resutls i.e. asking people what they think to try and triangulate results w data collection/coding/findings?}

\section{RQ1: People's Information Needs about COVID-19}

In RQ1, we summarise popular topics from Twitter to determine people's information requirements during the COVID-19 pandemic.

\subsection{Approach}
\label{sec:RQ1Approach}

\begin{table*}[]
\centering
\resizebox{0.9\textwidth}{!}{%
\begin{tabular}{lllll}\hline\hline
ID & Topics                   & Topic contribution                        & Keywords                                     & Theme \\
\hline\hline
1 &
  Transmission of COVID-19 & 0.1903 &
  care, health, worker, affect, cost &
  \multirow{5}{*}{\begin{tabular}[c]{@{}l@{}}\textbf{COVID-19}\\ 
  \textbf{information}\end{tabular}} \\
2  & Other viruses                & 0.1769                      & virus, call, spread, recover, scientist       &          \\
3  & Fake news                 & 0.1537                       & pandemic, live, realdonaldtrump, part, global &          \\
4  & The origin of COVID-19             & 0.1327               & happen, find, article, Chinese, cure          &          \\
5  & Daily COVID-19 data & 0.1708 & case, report, dead, positive, rate            &          \\
\hline
6 &
  Self-isolated & 0.1451 &
  hope, isolation, post, sense, mind &
  \multirow{7}{*}{\begin{tabular}[c]{@{}l@{}}\textbf{Protection}\\ \textbf{approaches}\end{tabular}} \\
7  & vaccine development              &     0.1776           & life, vaccine, fund, stop, risk               &          \\
8  & Stay at home                & 0.0969                     & lockdown, rest, move, fail, shut              &          \\
9  & Wear face masks           & 0.1377                    & doctor, mask, medical, face, wear             &          \\
10 & PPE shortage               & 0.1521                      & patient, hospital, fight, human, quarantine   &          \\
11 & Wash hands                & 0.1544               & home, check, safe, hand, play                 &          \\
12 & Antibody test                    & 0.1584                & test, person, symptom, infect, positive       &          \\
\hline
13 &
  Evidence of governments' response for COVID-19 & 0.1172 &
  time, crisis, news, learn, survive &
  \multirow{8}{*}{\begin{tabular}[c]{@{}l@{}}\textbf{Impact on}\\ \textbf{society}\end{tabular}} \\
14 & International travel            & 0.0959                 & country, start, hear, economic, flight        &          \\
15 & Aviation industry           & 0.1184           & today, media, travel, service, social         &          \\
16 & Government policies    & 0.1130  & state, stop, problem, force, lockdown         &          \\
17 &
  Impact on school/ university &  0.1409  &
  government, child, student, response, money &
   \\
18 & Protect domestic abuse survivors & 0.1440 & people, kill, year, sound, chance             &          \\
19 & Impact on women       & 0.1439          & work, public, result, high, strategy          &          \\
20 & Labour market         & 0.1909               & month, bill, protect, people, office          &          \\
\hline
21 &
  Small business & 0.1130 &
  support, economy, company, important, measure &
  \begin{tabular}[c]{@{}l@{}}\textbf{Impact on }\\ 
  \textbf{business}

  \end{tabular}
  \\ \hline\hline
\end{tabular}%
}
\caption{Topics generated from tweets}
\vspace{-6mm}
\label{tab:twitterTopics}
\end{table*}

\begin{figure}[h]
\centering 
\includegraphics[width=0.45\textwidth, height=1.4cm]{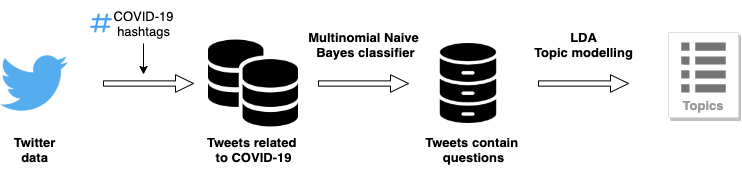} 
\vspace{-4mm}
\caption{Workflow of tweet analysis in RQ1} 
\label{Fig.approach}
\end{figure}

Fig~\ref{Fig.approach} shows the overall data collection and analysis process of RQ1.
First, we use the tweepy API~\cite{roesslein2009tweepy} to crawl all Twitter data from 22${^{nd}}$ March to 10${^{th}}$ June 2020.
We then create functions to iterate through timelines and filter out the ones that do not contain the hashtags we require and all the non-English tweets. 
As mentioned in Section~\ref{sec:hashtags}, we select the hashtags based on Breslin et al.'s research~\cite{breslin2020we} and our observation of popular COVID-19 related hashtags.

We noticed that tweets including questions often express users' requirements. 
For example, from ``\textit{Is that roughly one in 30 of the UK population has \#COVID19?}", we could observe that people want to know the proportion of British people diagnosed with COVID-19. % Ross et al.'s~\cite{ross1977structured} research also supports our observation. They used questions to define requirements and these serve as the basis for a development contract.
Therefore, we analysed tweets that contain questions. 
We applied a Multinomial Naive Bayes classifier~\cite{kibriya2004multinomial} to all the tweets data. 
The multinomial Naive Bayes classifier is a model which focuses on building word counts.
By creating a question word set to filter tweets involving those words and question marks, we can then tell whether the tweets are questions or not. 
The model used nps\_chats as training data and label two kinds of data, which are ``\textit{whQuestion}'' and ``\textit{ynQuestion}''.
``\textit{whQuestion}'' means questions that begin with ``\textit{wh}'', for example,``\textit{Why we need masks \#COVID-19?}''.
``\textit{ynQuestion}'' means questions that can be answered by yes or no, i.e, ``\textit{Is \#COVID-19 confirm cases number essential for public?}''.
% For example, ``\textit{Why we need masks??? \#COVID-19?}'' is a ``\textit{whQuestion}'' and ``\textit{Is \#COVID-19 confirm cases number essential for public?}'' is a ``\textit{ynQuestion}''.
% Furthermore, we applied this model to process our own data sets. 
% To get a more accurate analysis, we also drop retweets from our data set. 

%\chen{The current writing about the approach is too specific and engineering detailed. Please check Section 3.2 in my CSCW'18 paper \url{https://chunyang-chen.github.io/publication/proactiveEdit_cscw18.pdf} about how to write approach and empirical study results.}

% Because there existed many stopwords like ``so" or ``then" and they will occupy a large part of the text, which will become keywords if we not remove them.
%\chen{please check how to write this part by referring to my CSCW'18 paper }
We then apply text processing steps to all the tweets, including removing punctuation, lower-casing all characters, and excluding stop words. 
% We put some context-dependent words like ``\textit{COVID}'', ``\textit{coronavirus}'' to the stop words list. 
Finally, by adopting the Latent Dirichlet Allocation (LDA) model~\cite{vorontsov2015bigartm}, we extract the topic generated by our dataset with keywords and its importance, which name as topic contribution in Table~\ref{tab:twitterTopics}.
%To find the best model, we apply 10-fold cross validation.
%10-fold cross-validation will randomly divide the sample data into ten parts, randomly select nine parts as the training set, and the remaining part as the test set. 
%When a round of training is completed, randomly select nine copies to train the data. 
%After ten rounds, select the optimal model.

\subsection{Results}
We collect 56M tweets at first. After the initial processing (remove duplicated tweets and non-text tweets), about 48M tweets are remaining. 
Then, after filtering out the non-questioning and non-English tweets, 3.47M are remaining.
The LDA model generates 32 topics based on the remaining tweets, and we filtered and identified 21 topics associated with their keywords, which are listed in Table~\ref{tab:twitterTopics}. 
Finally, according to those topics, we extract them into four themes: COVID-19 information, protect approaches, related to society, and associated with the business.

\textbf{Theme 1: COVID-19 information}

The COVID-19 information theme contains five topics. 
The first topic is the transmission of COVID-19, which has a 0.19 topic contribution. 
Tweets in this topic are curious about how the virus spreads between individuals, whether it spreads between animals and humans, and the best way to stop it.
For example, ``\textit{Are schools driving \#COVID-19 transmission? What is the number?}". 
The second topic is the relation with other viruses, like ``\textit{Is COVID-19 has similar threats and severity with SARS and Ebola?}".
Nearly 83\%(373 posts) of tweets on this topic talk about scientists' research on past viruses. 
The third topic is fake news, which is also a hot topic in this area.
In March 2020, 582 fake news have been detected, such as researchers discovering a cure for COVID-19 and Israel has developed a vaccine for COVID-19~\cite{Anexplor99:online}. 
The next topic is about the origin of COVID-19, which has 13\% (783) tweets talking about it. 
People ask questions such as ``\textit{Where Coronavirus from?}"
In the last topic, people question about cases and death numbers, 82\% of tweets mention ``\textit{case number}, i.e. ``\textit{What is the case number today in...}", and about 45\% of posts involve ``\textit{How many people dead?...}". 
People question the real situation about COVID-19 to find out if they are safe or being protected.

%\chen{After reading the results, I cannot feel that you are analyzing questions? looks like normal tweets...}

\textbf{Theme 2: Protection approaches}

The second theme contains seven topics, which contain tweets talk about protections during the pandemic. 
Among them, self-protection and vaccine development are the two hottest topics. 
In this theme, 18\% of tweets discuss vaccine development like ``\textit{
how will the vaccine protect us against the disease?}". 
And 14\% of tweets talk about self-isolation such as ``\textit{Can I stay with families during \#COVID-19 self-isolation?}". 
About 13\% of tweets related to wear masks, and 15\% of tweets mention how to wash hands correctly. 
People want to avoid exposure to the virus, and they are eager to find a way to protect themselves. 

\textbf{Theme 3: Society issues}

The third theme is related to society. 
From the topics we summarise, people question the aviation industry, government responses, policies to handling the pandemic, and  the labor market. 
According to Maneenop et al.~\cite{maneenop2020impacts}, since the WHO declared the global pandemic outbreak, the aviation industry's cumulative negative abnormal return rate is 24.42\% within five days. 
Governments announced new policies, such as lockdown, to try and control the pandemic, and encouraging people to stay at home and self-isolate to avoid virus transmission.
% For example, the Victorian government of Australia proposed a work from home policy in August 2020. 
The pandemic has dramatically affected the labor market and the unemployment rate has increased sharply in many countries. 
From Bartik et al.'s statistics~\cite{bartik2020measuring}, in the week ending 14th March 2020, there were 250,000 initial unemployment insurance claims in the United States, about 20\% more than in the previous week. 
Therefore, people question how the epidemic affects their income and lives and ask for ways to help themselves.

\textbf{Theme 4: Related with business}

%\chen{I do not see people's questions about business. Can you please give more business questions that people are concerned with?}
The last theme is about business, especially small businesses. 
People  discuss concerns about small businesses on Twitter because many small businesses are financially vulnerable and overwhelmed during the epidemic.
For example, ``\textit{What does the Coronavirus Aid, Relief, and Economic Security (CARES) Act mean for business? \#covid19 \#coronavirus \#caresact \#smallbusiness}".
The example above asks about laws relate to small businesses during COVID-19. 
From the keywords in this theme, we can see that about 32\% of tweets included the word ``\textit{support}", people question the specific contents of financial support.  
% An example of this area is ``\textit{How can I access financial support? \#ExcludedUK \#seiss \#vaccination \#COVID19...}"
Our analyses reveal that people, especially small business owners, questioned the latest policies, laws, and financial support details.

\begin{center}
 \begin{tcolorbox}[colback=black!5!white,colframe=black!75!black,bottom=-0.05pt,top=-0.05pt]
By analyzing COVID-19 related tweets, we notice that people are concerned with daily situations (new cases/death/total cases) related to COVID-19, the ways to protect themselves in the pandemic, the effect of COVID-19 to their daily life, and COVID-19's impact on the economic market.
\end{tcolorbox}
\end{center}

\section{RQ2: Information Supply from COVID-19 Information Dashboard}

After exploring people's needs as expressed via Twitter during the COVID-19 pandemic, we wanted to know what kind of information is on the current dashboards. 
%Therefore, we will discuss the information supply in RQ2.

\subsection{Approach}

%\todo{Need somewhere to discuss why these dashboards chosen - perhaps in Motivation, Method or perhaps here??}

Based on the approach in Section~\ref{sec:open-coding}, we performed our data collection in the following way.
We analyzed the five dashboards in Table~\ref{tab:dashboards&codename} and categorized the information they were providing. 
Two groups of researchers from our team coded dashboards independently.
At first, each group recorded all data on dashboards and labeled them independently, such as ``\textit{total cases in Australia}" and ``\textit{test numbers in Boston}". 
Our groups created a table\footnote{Data release at URL: https://cutt.ly/Kjuzb5J} to store all the information and corresponding dashboards. 
Then, the researchers merged the information with the same label. 
For example, ``\textit{confirm cases in Australia}" are merged together with ``\textit{confirm cases in the United States}" as ``\textit{confirm cases in countries}". 
% Similarly, our groups merged data at the same level as well. 
% For instance, ``\textit{24 hours}" and ``\textit{7 days}" are merged as ``\textit{Time}" dimension. 
After the initial coding, the teams met and discussed any discrepancies until a consensus had been reached.

\subsection{Results}
%\chen{Please check Section 3.1 of my paper \url{https://arxiv.org/pdf/2008.06895.pdf} to see how I write something about taxonomy.}

\begin{figure}[htbp]
\centering 
\includegraphics[width=0.49\textwidth]{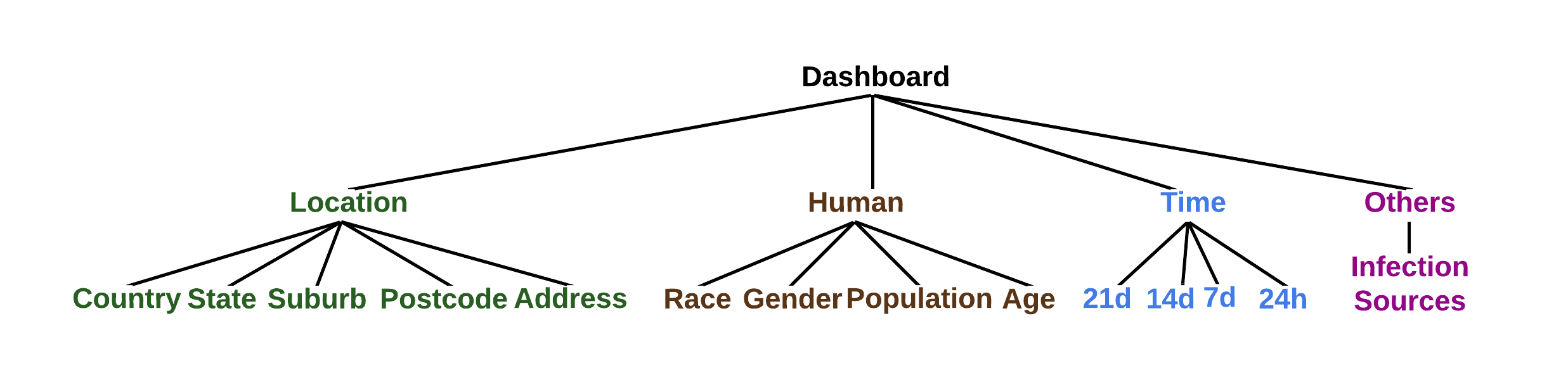} 
\vspace{-6mm}
\caption{Dashboard information taxonomy} 
\vspace{-4mm}
\label{Fig.taxnomy}
\end{figure}

% \begin{figure}[h] 
% \centering
% \includegraphics[width=0.4\textwidth]{figures/testinginfo.png} 
% \caption{Testing rates visualization in JHU} 
% \label{fig:testinginfo} 
% \end{figure}

We summarise four dimensions of information: location, human, time, and other, that are provided by the current COVID-19 dashboards, presented in Fig~\ref{Fig.taxnomy}. 
%These dimensions convey the information relates to COVID-19.
%And then, we determine whether the information is essential or is only specific to one country. 
We illustrate our summary of information on dashboards in Table~\ref{tab:Information on dashboards}. We use fractions to indicate the proportion of the five dashboards' information. 
For example, 1/5 means one of the five dashboards has this information. All the terms are referenced from the COVID Tracking project~\cite{DataDefi40:online}. 
% Active case number is the number of cases still considered to be infectious. Total case is the total number of confirmed plus probable cases of COVID-19. Confirmed cases are the total number of unique people with a positive PCR or other approved nucleic acid amplification test (NAAT). And to be asymptomatic means having coronavirus but not showing any symptoms at all.

\textbf{Location}

\begin{figure}
\centering
\subfigure[Special term on DXY dashboard]{\includegraphics[width=3.2cm]{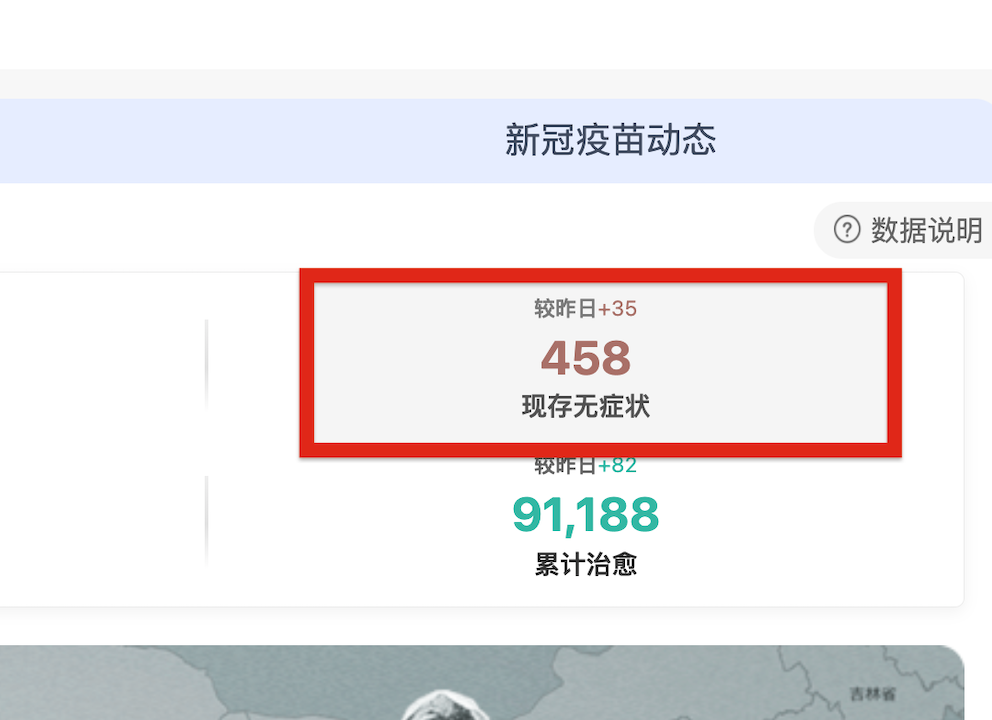}
\label{fig:Special term}}
\subfigure
[Address in case map on COVIDAu] 
{\includegraphics[width=3.2cm]{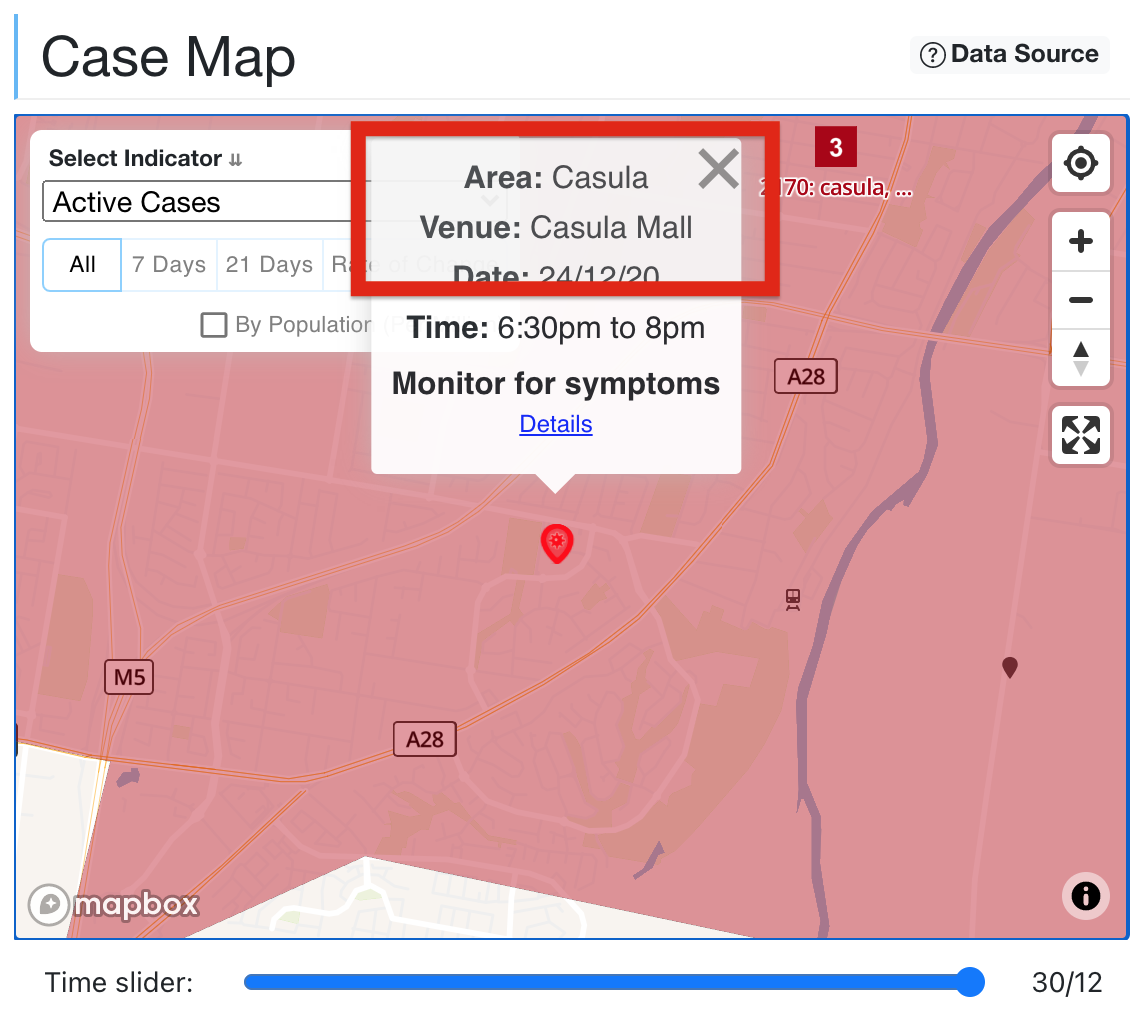}
\label{fig:Address in case map}} 
\caption{Explanation for location dimension} 
\vspace{-5mm}
\label{fig:details on dashboards_1} 
\end{figure}

In the location dimension, we identified five sub-dimensions. 
%To observe the information presented in those dashboards, we use fractions to show how many dashboards have this information.
Most of the dashboards focus on the country and states level. 
They implement data such as the total (5/5) and active cases (3/5) in each state or province.  
Moreover, there are certain features that only appear in one dashboard. 
For example, the term ``\textit{Asymptomatic infection cases}" on the DXY dashboard means patients with the COVID-19 in the incubation period are currently asymptomatic as they have no symptoms for the time being. 
This only shows on the Chinese version of the DXY dashboard in Fig~\ref{fig:Special term}. 
The city/postcode-level COVID-19 information (can be seen in Fig~\ref{fig:Address in case map}) from the COVIDAu dashboard, is also only identified in one dashboard. This may be due to that trustworthy data sources in different regions only provide limited types of data, so that it's hard for other dashboards to add those to their services. Besides, dashboards like the WHO dashboard, focus on serving the global information needs the most. They may not offer low level data to the public.  

%\chen{Note that most image caption in this paper are not detailed to describe the figure. Please revise!}

% Some information can seen from the data in Table~\ref{tab:Information on dashboards}. 
Three dashboards have test-related information on countries/states level and only one provides test location information on states level. 
Therefore, we can conclude that total, active cases, and death on countries and states level are the essential information supplies among the five dashboards.

\textbf{Human}

\begin{figure}
\centering
\subfigure[Population and total cases map on WHO]{\includegraphics[width=4cm]{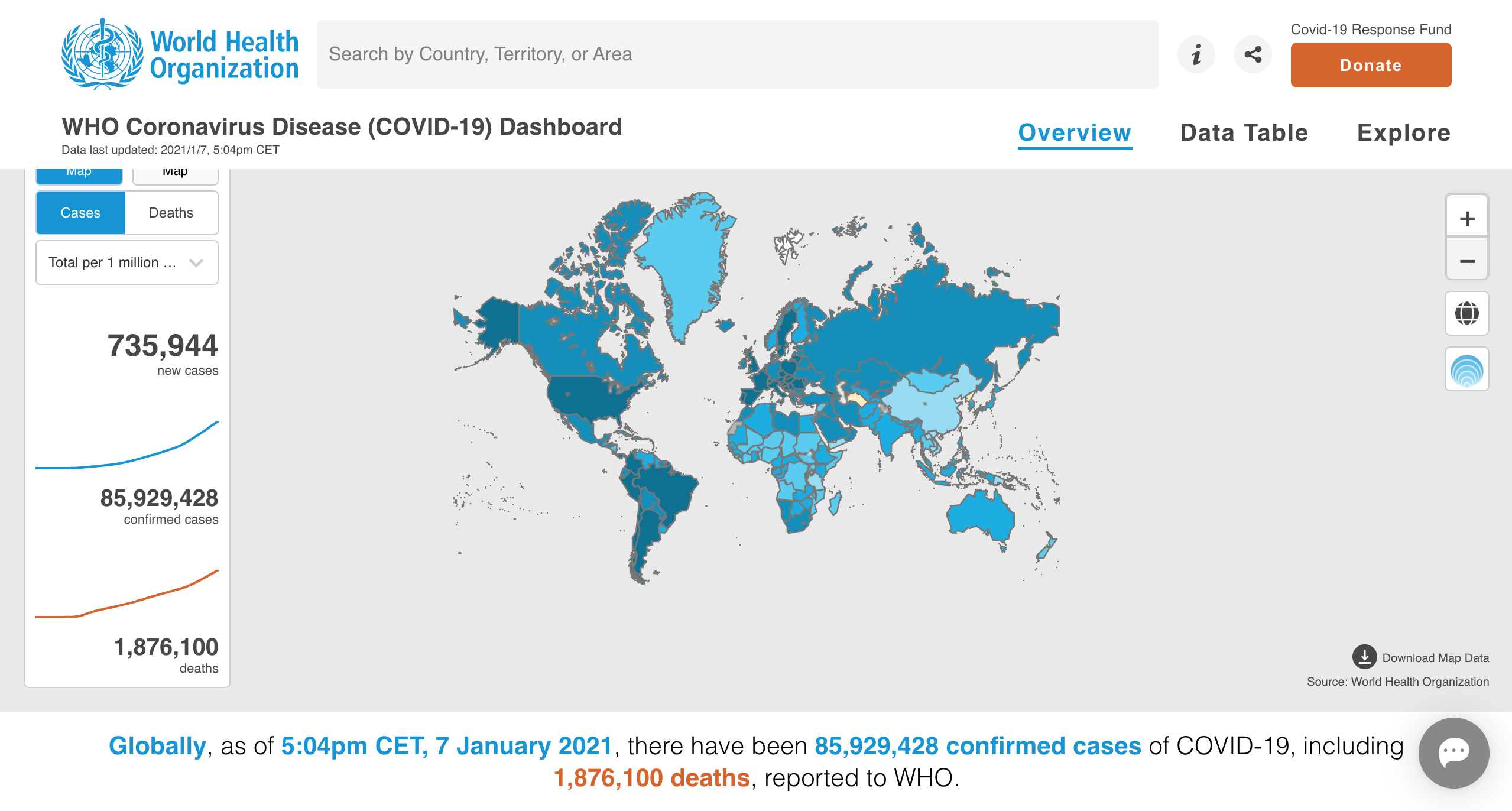}
\label{fig:Population and total cases map}
}
\subfigure[Gender visualisation on COVIDAu]{\includegraphics[width=4cm]{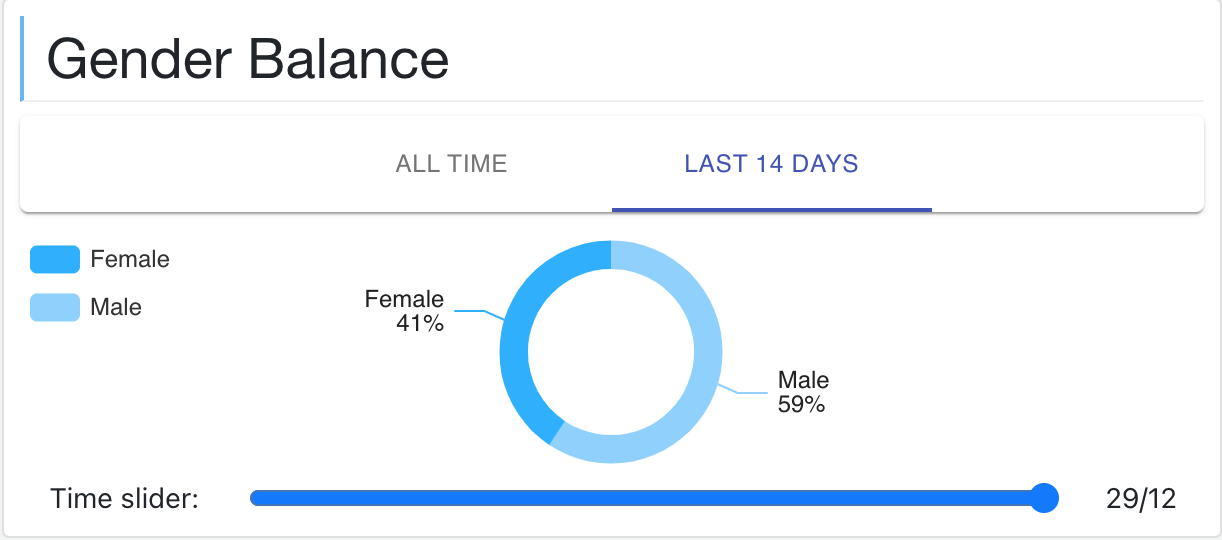}
\label{fig:Gender visualization}
}
\subfigure[Age distribution on COVIDAu]{\includegraphics[width=4cm]{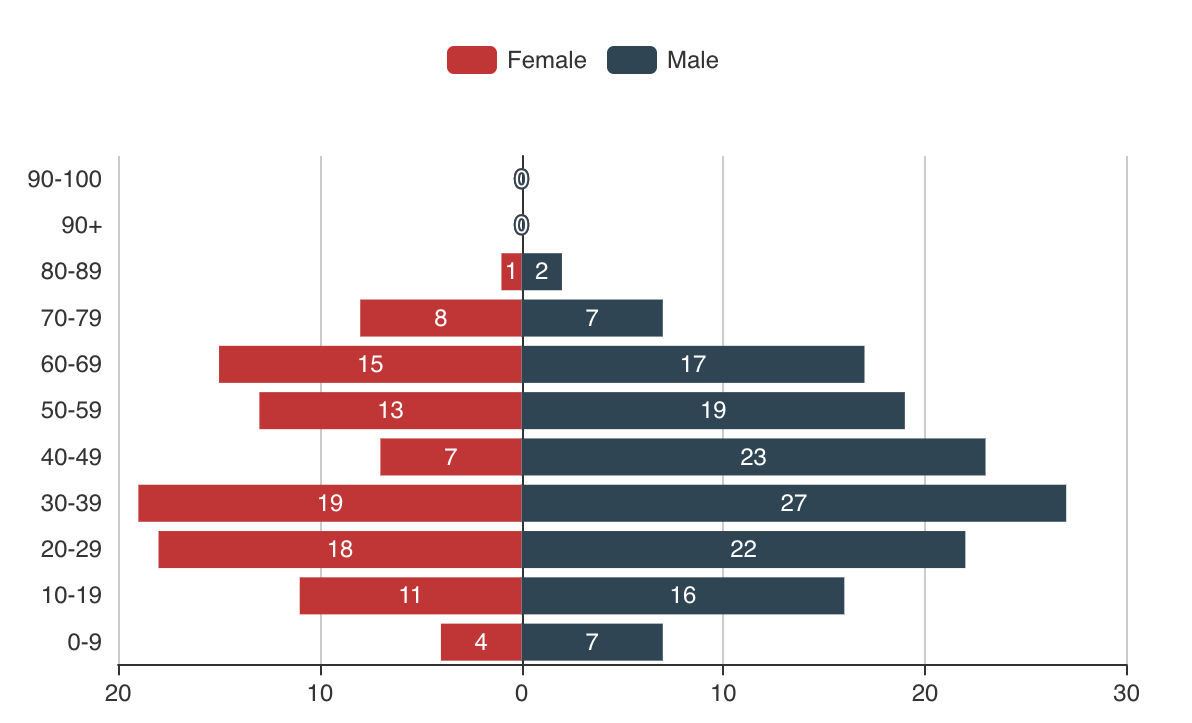}
\label{fig:Agedistribution}
}
\subfigure[Bubble map on WHO]{\includegraphics[width=4cm]{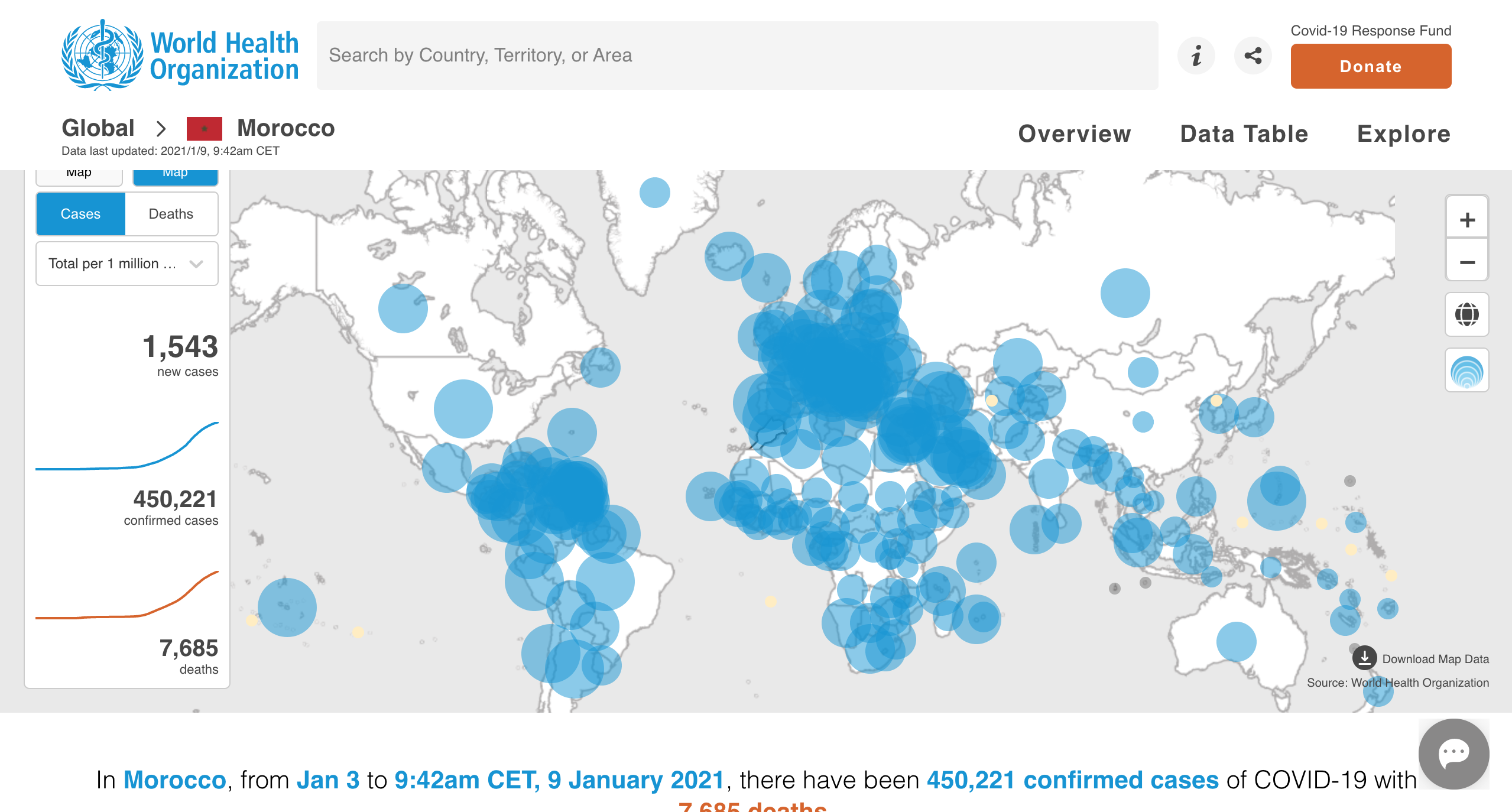}
\label{fig:Bubble chart on map}
}
\caption{Visualizations on COVID-19 dashboards} 
\vspace{-6mm}
\label{fig:details on dashboards_3} 
\end{figure}

We summarise four sub-dimensions for the human dimension. 
The first one is population, which occurs in four dashboards (4/5). All of them illustrate the relation between total case number and population.
For example, Fig~\ref{fig:Population and total cases map} is a screenshot from the WHO dashboard telling the relation between total cases number and population. 
Due to the diversity of the U.S. population, dashboard 1Point3Acres has a chart that shows the test results by race and ethnicity (1/5). 
Additionally, the dashboard COVIDAu presents the cases by age and gender (1/5), as shown in Fig~\ref{fig:Gender visualization} and Fig~\ref{fig:Agedistribution}. We could see that population is the one that is commonly used in all of the dashboards, whereas other factors like gender, age, and race are rarely seen. Since researchers~\cite{davies2020age, galasso2020gender} have tried to figure out the correlation between such data types and the infection, we may be able to see more visualizations on dashboards in the future. 
% Hence, the information of test results by race and ethnicity is country-specific. But this information has been deleted by the developers.

\textbf{Time}

The five dashboards indicate four different time intervals, which are 21-days, 14-days, 7-days, and 24-hours.
Four of these dashboards update total cases by 24 hours (4/5), which is essential information for the public.
And four out of five dashboards have the total death number by 7-days and 24h (4/5). 
However, some dashboards also implemented information like total cases with other periods, such as 14-days and 21-days. 

\textbf{Others}

The last factor is infection sources (5/5), which contains five categories: contracted overseas, unknown community transmission, contracted from confirmed cases, contracted interstates, and under investigation.

As shown in Table~\ref{tab:Information on dashboards}, it is clear that the existing dashboards focus on the case volumes, such as active, total, and death cases among countries and states.
We further divide the dashboards into two groups as shown in Table~\ref{tab:dashboards&codename}: the national dashboards and the global dashboards.  
We find that national dashboards like the COVIDAu and JHU dashboards focus on more specific information. 
For example, COVIDAu provides active cases at the postcode level or even exact address, convenient for local people to check whether their surroundings are safe. 
Both of the dashboards offer the mobility situations such as retail/recreation and public transport change, which can observe the COVID-19 impact on economic or people's daily life. 
They also pay more attention to human-related information.
% For instance, in the 1Point3Acre dashboard, the developers provide a line chart to show the tested number and total death number by race and ethnicity due to the diversity of the U.S. population. 
% Similarly, in the COVIDAu dashboard, developers analyze the gender and age impact on the COVID-19 cases. 
Besides, national dashboards tend to use different periods like 24 hours, 7 days, and a fortnight to explore the trend of COVID-19.

For global dashboards, developers provide cases and death tracking on a higher level: countries and states. We take WHO dashboard as an example.
Besides the country and state level data, the WHO dashboard also provides the relation between total cases or deaths and population. 
This dashboard update cases information every 24 hours and death data every seven days. 
Moreover, WHO pays attention to infection sources as well. 
All of the cases have been demonstrated how they transmit in society.

Based on the above analysis, we conclude that the global dashboard information is broader and more comprehensive because it is used to serve people around the world.
In contrast, a national dashboard provides more detailed information and pays more attention to exploring a specific country or region's epidemic data. 
The above phenomenon is because the national dashboard mainly serves local people, and they pay more attention to the situation in their area than global information.

\begin{center}
 \begin{tcolorbox}[colback=black!5!white,colframe=black!75!black,bottom=-0.05pt,top=-0.05pt]
We determine that current COVID-19 dashboard information focuses on the real\-time situation of COVID-19 at either country or state-level; human-related information focuses on the distribution of total cases by different ages; COVID-19 data for  different periods; and the infection sources of COVID-19.
\end{tcolorbox}
\end{center}

\section{RQ3: Visualization of COVID-19 Information}
After analysing information in current COVID-19 information dashboards, we noticed that different dashboards use various ways to visualize the same data type. 
In RQ3, we try to determine the different visualization approaches that the dashboards use for same data types.

\subsection{Approach}
Similar to the approach in Section~\ref{sec:open-coding}, we analyse the same five dashboards but concentrate on COVID-19 data visualizations and interactions. We had two groups from our research team perform open coding on the dashboards. 
% Please add the following required packages to your document preamble:
% \usepackage{multirow}
% \usepackage{graphicx}
% \usepackage{lscape}
\begin{landscape}
\begin{table}[]
\centering
\resizebox{1.35\textwidth}{!}{
\begin{tabular}{c|l|cccll|llll|clcl|l}
\hline\hline
\multicolumn{1}{l}{} &
   &
  \multicolumn{14}{c}{Dimensions} \\\hline\hline
\multicolumn{1}{l}{} &
   &
  \multicolumn{5}{c|}{Location} &
  \multicolumn{4}{c|}{Human} &
  \multicolumn{4}{c|}{Time} &
  \multicolumn{1}{c}{Others} \\
\multicolumn{1}{l}{} &
   &
  Countries &
  States &
  \begin{tabular}[c]{@{}c@{}}Suburb/\\  districts\end{tabular} &
  \multicolumn{1}{c}{Postcode} &
  \multicolumn{1}{c|}{Address} &
  \multicolumn{1}{c}{Race} &
  \multicolumn{1}{c}{Gender} &
  \multicolumn{1}{c}{Population} &
  \multicolumn{1}{c|}{Age} &
  21d &
  \multicolumn{1}{c}{14d} &
  7d &
  \multicolumn{1}{c|}{24h} &
  \multicolumn{1}{c}{Infection Sources} \\\hline
\multirow{3}{*}{Case} &

  Active cases &
  \begin{tabular}[c]{@{}c@{}}4/5\\ {\color{violet}(2M, 1L, 1D,}\\{ \color{violet}3T, 1Bu)}\end{tabular} &
  \begin{tabular}[c]{@{}c@{}}{\color{orange}3/5}\\ (2M, 1T, 1L)\end{tabular} &
  \begin{tabular}[c]{@{}c@{}}{\color{orange}1/5}\\ (1M)\end{tabular} &
  \multicolumn{1}{c}{\begin{tabular}[c]{@{}c@{}}{\color{orange}1/5}\\ (1M)\end{tabular}} &
  \multicolumn{1}{c|}{\begin{tabular}[c]{@{}c@{}}{\color{orange}1/5}\\ (1M)\end{tabular}} &
   &
   &\begin{tabular}[c]{@{}c@{}}1/5\\ (1M)\end{tabular}
   & 
   &
  \begin{tabular}[c]{@{}c@{}}1/5\\ (1M)\end{tabular} &
  \multicolumn{1}{c}{} &
  \begin{tabular}[c]{@{}c@{}}2/5\\ (1M, 1L)\end{tabular} &
  \begin{tabular}[c]{@{}c@{}}2/5\\ {\color{violet}(2L)}\end{tabular}
   & 
   \\
 &
  Total cases &
  \begin{tabular}[c]{@{}c@{}}{\color{orange}5/5}\\{\color{violet} ((3M, 2T, 2L,1D,}\\{ \color{violet} 
2Bu, 2H)}\end{tabular} &
  \begin{tabular}[c]{@{}c@{}}{\color{orange}5/5}\\ (3M,1H,1L,\\1T)\end{tabular} &
  \begin{tabular}[c]{@{}c@{}}{\color{orange}1/5}\\ (1M)\end{tabular} &
  \multicolumn{1}{c}{\begin{tabular}[c]{@{}c@{}}{\color{orange}1/5}\\ (1M)\end{tabular}} &
  \multicolumn{1}{c|}{\begin{tabular}[c]{@{}c@{}}1/5\\ (1M)\end{tabular}} &
   &
  \multicolumn{1}{c}{\begin{tabular}[c]{@{}c@{}}{\color{orange}1/5}\\ {\color{violet}(1D)}\end{tabular}} &
  \multicolumn{1}{c}{\begin{tabular}[c]{@{}c@{}}4/5\\ (3M,1T,1Bu)\end{tabular}} &
  \multicolumn{1}{c|}{\begin{tabular}[c]{@{}c@{}}{\color{orange}1/5}\\ {\color{violet}(1B)}\end{tabular}} &
  \begin{tabular}[c]{@{}c@{}}1/5\\ (1M)\end{tabular} &
  \multicolumn{1}{c}{\begin{tabular}[c]{@{}c@{}}1/5\\ (1T)\end{tabular}} &
  \begin{tabular}[c]{@{}c@{}}3/5\\ (2M,2H,\\1Bu)\end{tabular} &
  \multicolumn{1}{c|}{\begin{tabular}[c]{@{}c@{}}{\color{orange}4/5}\\ {\color{violet}(3H,2L,1M,}\\{ \color{violet}1Bu)}\end{tabular}} &
  \multicolumn{1}{c}{\begin{tabular}[c]{@{}c@{}}{\color{violet}5/5}\\ (1L, 2M, 2T)\end{tabular}} \\
 &
 
  Asymptomatic infection cases &
  \multicolumn{1}{l}{} &
  \begin{tabular}[c]{@{}c@{}}{\color{orange}1/5}\\ (1T)\end{tabular} &
  \multicolumn{1}{l}{} &
   &
   &
   &
   &
  \multicolumn{1}{c}{} &
   &
  \multicolumn{1}{l}{} &
   &
  \multicolumn{1}{l}{} &
  \multicolumn{1}{c|}{\begin{tabular}[c]{@{}c@{}}1/5\\ (1T)\end{tabular}} &
  \multicolumn{1}{c}{}
 \\\hline
   
Recovered &

  Total recovered &
 
  \begin{tabular}[c]{@{}c@{}}4/5\\ (1M, 4T)\end{tabular} &
  \begin{tabular}[c]{@{}c@{}}4/5\\ (1M, 1L, 2T)\end{tabular} &
  \begin{tabular}[c]{@{}c@{}}1/5\\ (1M)\end{tabular} &
   &
  \multicolumn{1}{c|}{\begin{tabular}[c]{@{}c@{}}1/5\\ (1M)\end{tabular}} &
   &
   &\multicolumn{1}{c}{\begin{tabular}[c]{@{}c@{}}1/5\\ (1M)\end{tabular}}
   &
   &
  \begin{tabular}[c]{@{}c@{}}1/5\\ (1M)\end{tabular} &
  \multicolumn{1}{c}{} &
  \begin{tabular}[c]{@{}c@{}}2/5\\ (2M)\end{tabular} &
   \begin{tabular}[c]{@{}c@{}}1/5\\ (1L)\end{tabular}&
   \\\hline
\multirow{6}{*}{Tested} &

  Tested &
  \begin{tabular}[c]{@{}c@{}}{\color{orange}3/5}\\ (2M, 1L, 1T,\\ 1Bu)\end{tabular} &
  \begin{tabular}[c]{@{}c@{}}{\color{orange}3/5}\\ (2M, 2T, 1L,\\ 1B)\end{tabular} &
  \multicolumn{1}{l}{} &
   &
   &
  \multicolumn{1}{c}{\begin{tabular}[c]{@{}c@{}}{\color{orange}1/5}\\ {\color{violet}(1B)}\end{tabular}} &
   &\multicolumn{1}{c}{\begin{tabular}[c]{@{}c@{}}1/5\\ (1M)\end{tabular}}
   &
   &
  \begin{tabular}[c]{@{}c@{}}1/5\\ (1M)\end{tabular} &
   &
  \begin{tabular}[c]{@{}c@{}}3/5\\ (1M,1L,1T)\end{tabular} &
  \multicolumn{1}{c|}{\begin{tabular}[c]{@{}c@{}}2/5\\ (2L,1T)\end{tabular}} &
   \\
 &

  Positive(\%) &
  
  \multicolumn{1}{c}{\begin{tabular}[c]{@{}c@{}}1/5\\ (1T)\end{tabular}} &
  \multicolumn{1}{c}{\begin{tabular}[c]{@{}c@{}}1/5\\ (1T)\end{tabular}} &
  \multicolumn{1}{l}{} &
   &
   &
  \multicolumn{1}{c}{\begin{tabular}[c]{@{}c@{}}{\color{orange}1/5}\\ (1B)\end{tabular}} &
   &
   &
   &
  \multicolumn{1}{l}{} &
    &
   \multicolumn{1}{c}{\begin{tabular}[c]{@{}c@{}}1/5\\ (1T)\end{tabular}}&
   &
   \\
 &
 
  Positive tests &
  \multicolumn{1}{c}{\begin{tabular}[c]{@{}c@{}}1/5\\ (1H)\end{tabular}} &
  \multicolumn{1}{c}{\begin{tabular}[c]{@{}c@{}}1/5\\ (1B, 1H)\end{tabular}} &
  \multicolumn{1}{l}{} &
   &
   &
  \multicolumn{1}{c}{\begin{tabular}[c]{@{}c@{}}{\color{orange}1/5}\\ (1B)\end{tabular}} &
   &
   &
   &
  \multicolumn{1}{l}{} &
   &
  \multicolumn{1}{c}{\begin{tabular}[c]{@{}c@{}}1/5\\ (1H)\end{tabular}} &
  \multicolumn{1}{c|}{\begin{tabular}[c]{@{}c@{}}2/5\\ (2H)\end{tabular}} &
   \\
 &

  Negative tests &
  \multicolumn{1}{c}{\begin{tabular}[c]{@{}c@{}}1/5\\ (1H)\end{tabular}} &
  \multicolumn{1}{c}{\begin{tabular}[c]{@{}c@{}}1/5\\ (1B,1H)\end{tabular}} &
  \multicolumn{1}{l}{} &
   &
   &
  \multicolumn{1}{c}{\begin{tabular}[c]{@{}c@{}}{\color{orange}1/5}\\ (1B)\end{tabular}} &
   &
   &
   &
   &
   &
  \multicolumn{1}{c}{\begin{tabular}[c]{@{}c@{}}1/5\\ (1H)\end{tabular}} &
   \multicolumn{1}{c|}{\begin{tabular}[c]{@{}c@{}}1/5\\ (1H)\end{tabular}}&
   \\
 &
  
  Pending test &
  \multicolumn{1}{l}{} &
  \multicolumn{1}{l}{} &
  \multicolumn{1}{l}{} &
   &
   &
   &
   &
   &
   &
  \multicolumn{1}{l}{} &
   &
  \multicolumn{1}{l}{} &
   &
   \\
 &
  Test locations &
  \multicolumn{1}{l}{} &
  \begin{tabular}[c]{@{}c@{}}1/5\\ (1T)\end{tabular} &
  \multicolumn{1}{l}{} &
   &
   &
   &
   &
   &
   &
  \multicolumn{1}{l}{} &
   &
  \multicolumn{1}{l}{} &
   &
   \\\hline
    
\multirow{2}{*}{Death} &

  Total death &
  \begin{tabular}[c]{@{}c@{}}5/5\\ {\color{violet}(2M, 1D, 3L,}\\{ \color{violet} 2H, 4T, 1Bu)}\end{tabular} &
  \begin{tabular}[c]{@{}c@{}}5/5\\ {\color{violet}(3M,3T,2L,}\\{ \color{violet}2H)}\end{tabular} &
  \begin{tabular}[c]{@{}c@{}}1/5\\ (1M)\end{tabular} &
   &
   &
  \multicolumn{1}{c}{\begin{tabular}[c]{@{}c@{}}1/5\\{\color{violet} (1B)}\end{tabular}} &
   &
  \multicolumn{1}{c}{\begin{tabular}[c]{@{}c@{}}3/5\\ (2M, 1Bu)\end{tabular}} &
   &
  \begin{tabular}[c]{@{}c@{}}1/5\\ (1M)\end{tabular} &
  \multicolumn{1}{c}{} &
  \begin{tabular}[c]{@{}c@{}}{\color{orange}4/5}\\ (2M,2L,2H,\\1Bu)\end{tabular} &
   \begin{tabular}[c]{@{}c@{}}{\color{orange}4/5}\\ (1L,3H,1M,\\1Bu)\end{tabular}&
   \\
 &
  Fatality &
  \begin{tabular}[c]{@{}c@{}}3/5\\ (1T, 1M, 1Bu)\end{tabular} &
  \begin{tabular}[c]{@{}c@{}}2/5\\ (2H)\end{tabular} &
   &
   &
   &
   &
   &
   &
   &
  \multicolumn{1}{l}{} &
   &
   &
   &
   \\\hline

\multirow{3}{*}{Data from hospitals} &

  In hospital &
  \begin{tabular}[c]{@{}c@{}}2/5\\ ((1T, 1M))\end{tabular} &
  \begin{tabular}[c]{@{}c@{}}2/5\\ (1M, 1T, 2L)\end{tabular} &
  \begin{tabular}[c]{@{}c@{}}1/5\\ (1M)\end{tabular} &
   &
   &
   &
   &
   &
   &
  \begin{tabular}[c]{@{}c@{}}1/5\\ (1M)\end{tabular} &
   &
  \begin{tabular}[c]{@{}c@{}}2/5\\ (1M,1T)\end{tabular} &
   \begin{tabular}[c]{@{}c@{}}2/5\\ (1L,1T)\end{tabular}&
   \\
 &

  In ICU w/ Ventilator &
  \begin{tabular}[c]{@{}c@{}}1/5\\ (1M)\end{tabular} &
  \begin{tabular}[c]{@{}c@{}}1/5\\ (1M)\end{tabular} &
  \begin{tabular}[c]{@{}c@{}}1/5\\ (1M)\end{tabular} &
   &
   &
   &
   &
   &
   &
  \begin{tabular}[c]{@{}c@{}}1/5\\ (1M)\end{tabular} &
   &
  \begin{tabular}[c]{@{}c@{}}1/5\\ (1M)\end{tabular} &
   &
   \\
 &
  
  In ICU &
  \begin{tabular}[c]{@{}c@{}}1/5\\ (1M)\end{tabular} &
  \begin{tabular}[c]{@{}c@{}}1/5\\(1M,1T,1L)\end{tabular} &
  \begin{tabular}[c]{@{}c@{}}1/5\\ (1M)\end{tabular} &
   &
   &
   &
   &
   &
   &
  \begin{tabular}[c]{@{}c@{}}1/5\\ (1M)\end{tabular} &
   &
  \begin{tabular}[c]{@{}c@{}}1/5\\ (1M)\end{tabular} &
  \begin{tabular}[c]{@{}c@{}}1/5\\ (1L)\end{tabular} &
   \\\hline
\multirow{5}{*}{Mobility} &
  Retail/ Recreation change &
  \begin{tabular}[c]{@{}c@{}}1/5\\ (1M)\end{tabular} &
  \begin{tabular}[c]{@{}c@{}}2/5\\ (1M, 1L)\end{tabular} &
  \begin{tabular}[c]{@{}c@{}}1/5\\ (1M)\end{tabular} &
   &
   &
   &
   &
   &
   &
  \begin{tabular}[c]{@{}c@{}}1/5\\ (1M)\end{tabular} &
   &
  \begin{tabular}[c]{@{}c@{}}1/5\\ (1M)\end{tabular} &
   &
   \\
 &
  
  Supermarket/ pharmacy change &
  \begin{tabular}[c]{@{}c@{}}1/5\\ (1M)\end{tabular} &
  \begin{tabular}[c]{@{}c@{}}2/5\\ (1M, 1L)\end{tabular} &
  \begin{tabular}[c]{@{}c@{}}1/5\\ (1M)\end{tabular} &
   &
   &
   &
   &
   &
   &
  \begin{tabular}[c]{@{}c@{}}1/5\\ (1M)\end{tabular} &
   &
  \begin{tabular}[c]{@{}c@{}}1/5\\ (1M)\end{tabular} &
   &
   \\
 &
  Parks change &
  \begin{tabular}[c]{@{}c@{}}1/5\\ (1M)\end{tabular} &
  \begin{tabular}[c]{@{}c@{}}2/5\\ (1M, 1L)\end{tabular} &
  \begin{tabular}[c]{@{}c@{}}1/5\\ (1M)\end{tabular} &
   &
   &
   &
   &
   &
   &
  \begin{tabular}[c]{@{}c@{}}1/5\\ (1M)\end{tabular} &
   &
  \begin{tabular}[c]{@{}c@{}}1/5\\ (1M)\end{tabular} &
   &
   \\
 &
  
  Residential change &
  \begin{tabular}[c]{@{}c@{}}1/5\\ (1M)\end{tabular} &
  \begin{tabular}[c]{@{}c@{}}2/5\\ (1M, 1L)\end{tabular} &
  \begin{tabular}[c]{@{}c@{}}1/5\\ (1M)\end{tabular} &
   &
   &
   &
   &
   &
   &
  \begin{tabular}[c]{@{}c@{}}1/5\\ (1M)\end{tabular} &
   &
  \begin{tabular}[c]{@{}c@{}}1/5\\ (1M)\end{tabular} &
   &
   \\
 &

  Public transport &
  \begin{tabular}[c]{@{}c@{}}1/5\\ (1M)\end{tabular} &
  \begin{tabular}[c]{@{}c@{}}2/5\\ (1M, 1L)\end{tabular} &
  \begin{tabular}[c]{@{}c@{}}1/5\\ (1M)\end{tabular} &
   &
   &
   &
   &
   &
   &
  \begin{tabular}[c]{@{}c@{}}1/5\\ (1M)\end{tabular} &
   &
  \begin{tabular}[c]{@{}c@{}}1/5\\ (1M)\end{tabular} &
   &
   \\\hline
\multicolumn{1}{l|}{Additional info.} &
 
  Employment &
  \multicolumn{1}{l}{} &
  \begin{tabular}[c]{@{}c@{}}2/5\\ (1M, 1L)\end{tabular} &
  \begin{tabular}[c]{@{}c@{}}1/5\\ (1M)\end{tabular} &
   &
   &
   &
   &
   &
   &
  \begin{tabular}[c]{@{}c@{}}1/5\\ (1M)\end{tabular} &
   &
  \begin{tabular}[c]{@{}c@{}}1/5\\ (1M)\end{tabular} &
   & 
   \\\hline\hline
\end{tabular}
}

\caption{Information on dashboards. Code for different visualizations Tree map:Tr, Map: M, Linear chart: L, Dount chart: D, Bar chart: B, Histogram: H, Tabular data: T, Bubble chart: Bu}
\label{tab:Information on dashboards}
\end{table}
\end{landscape}

% Besides, if we apply a higher zoom level, there are also related data in specific postcodes and Highly epidemic areas. 

\begin{table}[!htbp]
\centering
\begin{tabular}{ccc}
\hline\hline
Interaction techniques & Explanation                           \\
\hline\hline
Select  &  Mark and track of them       \\
Explore &  Examine subsets data from large data sets \\
Reconfigure & Eearrange data/ adjust baseline/ move items \\
Encode & Change how the data represent \\
Abstract/ Elaborate  & Choose the level of data representation \\
\multirow{2}{*}{Connect} &  See relationships between data items \\ & by clicking or hovering cursor \\
Others & Undo/Redo/Change configurations/... \\
\hline\hline
\end{tabular}
\caption{Interaction techniques and its explanation}
\vspace{-5mm}
\label{tab:Interaction techniques and its explanation}
\end{table}

% \revision{We applied code names for different charts in the following sections: .}

% \begin{table}[!htbp]
% \centering
% \begin{tabular}{ccc}
% \hline\hline
% Chart name & Code name                           \\
% \hline\hline
% Tree map  & Tr        \\
% Map & M \\
% Linear chart & L \\
% Donut chart & D \\
% Bar chart & B \\
% Histogram & H \\
% Tabular data & T \\
% Bubble chart & Bu \\
% \hline\hline
% \end{tabular}
% \caption{Chart names and code names}
% \vspace{-5mm}
% \label{tab:Chart names and code names}
% \end{table}

%\todo{Need more details on analysis techniques used here?

%Including who did coding, how many, more details on the cross-validation etc?

%}
\noindent
The members are masters students with computer science backgrounds.
Each team manually recorded and categorized all the charts on dashboards and labeled them with categories. 
Our teams then aggregated the them into eight types, such as bubble charts and bar charts.
After finishing the coding, our researchers met and discussed any discrepancies until consensus was reached. 
Furthermore, by using Yi et al.'s criteria ~\cite{interactionTypes} on interaction types, we matched all visualizations to the seven interaction techniques in Table~\ref{tab:Interaction techniques and its explanation}.

%\todo{Are there dashboard reviews that can be analysed like the twitter messages to determine (i) strenghs, (ii) weakneses and (iii) gaps from end user perspective - to compare to the analysis done below???}

% Please add the following required packages to your document preamble:
% \usepackage{multirow}
% \usepackage{graphicx}
% \usepackage{lscape}
\begin{table*}[]
\centering
\resizebox{0.85\textwidth}{!}{%
\begin{tabular}{clccccclcc}
\hline\hline
\multicolumn{1}{l}{}                                                               &                    & Map                                                                                              & Bubble map           & Tableau data                                                                                                                                                                           & Histogram            & Donut chart          & \multicolumn{1}{c}{Treemap} & Line chart           & Bar chart            \\
\hline\hline
\multirow{8}{*}{\begin{tabular}[c]{@{}c@{}}interaction\\  techniques\end{tabular}} & Select             & 1,2,3                                                                                            &                      & 3                                                                                                                                                                                      & \multicolumn{1}{l}{} & \multicolumn{1}{l}{} &                             & \multicolumn{1}{l}{} & \multicolumn{1}{l}{} \\
                                                                                   & Explore            & 1,2,3                                                                                            & 3,5                    & 1,2,4                                                                                                                                                                                  & 1,2,5                & 1                    & \multicolumn{1}{c}{1}       & \textbf{1,2,3,4}             & 1,2                  \\
                                                                                   & Reconfigure        & \textbf{1,2,3,4}                                                                                          & 3,5                    & 2,3                                                                                                                                                                                    & \textbf{1,2,3,5}             & 1                    & \multicolumn{1}{c}{1}       & \textbf{1,2,3,4}              & 1,2                  \\
                                                                                   & Encode             & \multicolumn{1}{l}{}                                                                             & \multicolumn{1}{l}{} & \multicolumn{1}{l}{}                                                                                                                                                                   & \multicolumn{1}{l}{} & \multicolumn{1}{l}{} &                             & \multicolumn{1}{l}{} & \multicolumn{1}{l}{} \\
                                                                                   & Abstract/Elaborate & 1,2,3                                                                                            & 3,5                    & 1,2                                                                                                                                                                                    & 1,2,5                & \multicolumn{1}{l}{} &                             & 1                    & \multicolumn{1}{l}{} \\
                                                                                   & Filter             & 1                                                                                                & 3,5                    & \multicolumn{1}{l}{}                                                                                                                                                                   & 1,2                  & 1,2                  &                             & 1,2                  & 1,2                  \\
                                                                                   & Connect            & \textbf{1,2,3,4}                                                                                          & 3,5                    & 2,3,4                                                                                                                                                                                  & \textbf{1,2,3,5}              & 1,2                  &                             & \textbf{1,2,3,5}              & 1,2                  \\
                                                                
\hline\hline
\end{tabular}%
}
\caption{Visualization and interaction (1: COVIDAu, 2: 1Point3Acres, 3: JHU, 4: DXY, 5: WHO) }
\label{tab: data visualization and interaction}
\vspace{-5mm}
\end{table*}

\subsection{Results}

After conducting the experiments, we summarised the data and visualizations used by the selected dashboards.
We applied the same taxonomy Fig~\ref{Fig.taxnomy} that is used in RQ2 and summarised eight visualization approaches. 
To simplify the content in Table~\ref{tab:Information on dashboards}, we use abbreviations to represent different visualization methods.
In Table~\ref{tab:Information on dashboards}, we can observe which visualization method is more suitable for what kind of data. 
We also summarise the standard interaction techniques implemented in different kinds of visualization.

\textbf{Data and visualization}

As shown in Table~\ref{tab:Information on dashboards}, we found that a map visualisation is commonly applied (74\%) when presenting the data related to the location dimension.
Sometimes, it is also combined with a bubble chart, that can present COVID-19 data in each part of the world, like Fig~\ref{fig:Bubble chart on map}. 
Line charts are also used to present trends, such as active cases in 7\-days in the United States.
Developers implement tabular data to show the active and all cases of an area so that users can understand detailed epidemic data.
Only one dashboard uses donuts charts to present the percentages of active cases, total cases, and total death in the top five countries and states, which compare the proportion of cases in different countries or regions. 
We also found that histograms are frequently used to show total death distributions in states (5/5).

\begin{figure}
\centering
\subfigure[Relation between population and confirm cases]{\includegraphics[width=3.5cm]{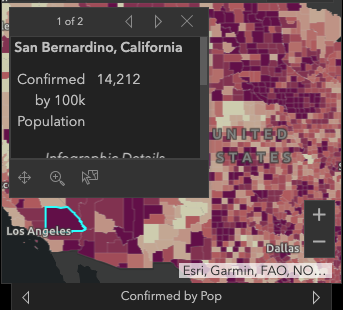}
\label{fig:population}}
\subfigure
[Multi-dimensional data on map] 
{\includegraphics[width=5cm]{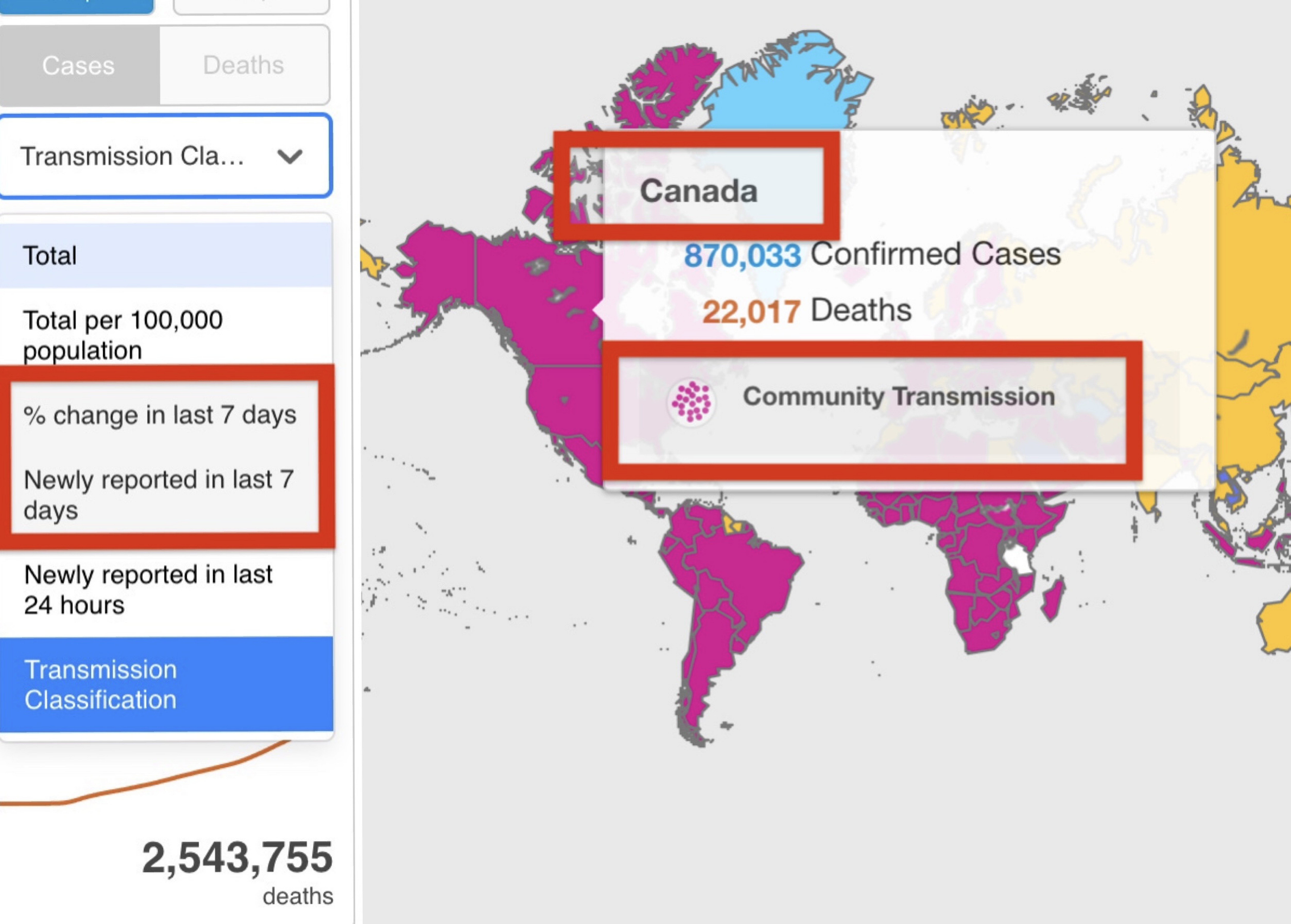}
\label{fig:drop-box}} 
\caption{Examples for data and visualization} 
\vspace{-5mm}
\end{figure}

The five dashboards we analysed have fewer visualizations on the human dimension than the other three dimensions.
Three out of five dashboards explore the relationship between population and the number of confirmed cases, the total number of deaths by applying a map, as can be seen in Fig~\ref{fig:population}. Such visualizations can effectively compare the infection rate per population in different areas rather than simple numbers. During the development of the COVIDAu dashboard, the authors have received certain requests for this type of visualization from users.

\begin{figure}[h]
\centering
\includegraphics[width=0.4\textwidth]{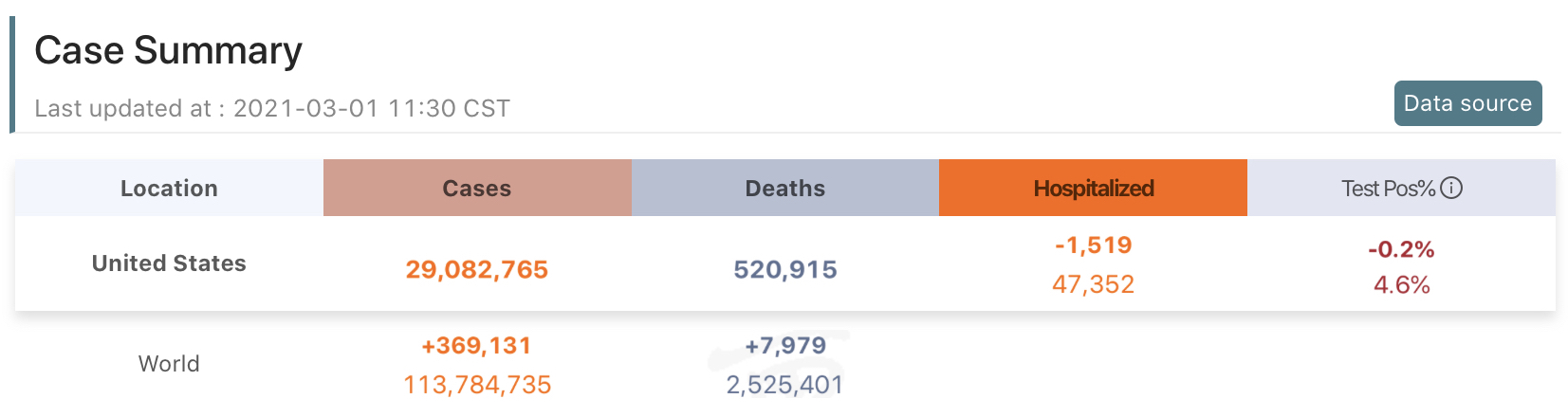} 
\vspace{-3mm}
\caption{Daily new cases in tabular data} 
\vspace{-3mm}
\label{fig:dailydata} 
\end{figure}

Concerning the time dimension, all of the dashboards we analysed use a line chart (2/2) to present the trends of active cases in 24-hours.
And three of them use histogram (3/5) to present the distribution of total cases in 24-hours. 
From the tabular data in Fig~\ref{fig:dailydata}, users can directly see the changes in daily data. 
For example, ``\textit{+861,748}'' means there are 861,748 new cases within 24 hours. 
A histogram demonstrates the number of cases changing.
% There is also one dashboard showing confirmed cases in 7 days by map and bubble chart.
% Users can observe the areas of each bubble to understand information like total cases and deaths number between different countries. 

In others dimension, all of the dashboards analyze the infection sources (5/5), but they each apply different visualizations to show this data. 
One of them uses line chart (1/5) to show the number of cases caused by each infection source. 
Two dashboards apply tabular data (2/5), and another two implement map (2/5). 
By looking at the map, users can understand the infection situation around their location and remind users of severe infections in the community to protect their safety. 
Furthermore, the tabular data allows users to understand the cases caused by each source of infection.

We observed that one visualization type can often be used to illustrate different types of data dimensions.
We divided the charts into three categories based on the number of data dimensions: one-dimensional data, two-dimensional data, and multi-dimensional data.
According to our analysis, developers use map visualization to present time, location, or other information, which makes it belongs to the multi-dimensional category.
For instance, in Fig~\ref{fig:drop-box}, developers apply a drop-down box for users to choose data dimensions.
Line charts and histograms typically contain two-dimensional data.
They provide time dimension and one the other dimension (e.g., human dimension/ other dimension) to show the trends and distributions. Such multi-dimensional data visualizations enrich the information users can acquire from charts, and can help users better understand the tendency of the virus from different aspects. 

One special data visualization used is on the COVIDAu dashboard, as shown in Fig~\ref{fig:Agedistribution}.
It displays the total number of cases of different genders and ages. 
This dashboard also applies a donut chart in Fig~\ref{fig:Gender visualization} to compare the total cases in male, female, and unknown sex, showing the percentage of different genders. 
%A bar chart has shown the total cases in different generations.
%Similar to the age visualization, bar charts have also been implemented to explore the tested results and total deaths in various races in the U.S.

\textbf{Visualization and interaction}

\begin{figure}
\centering 
\subfigure[Before reconfigure interaction]{
\label{Fig.sub.1}
\includegraphics[width=0.35\textwidth, height=2cm]{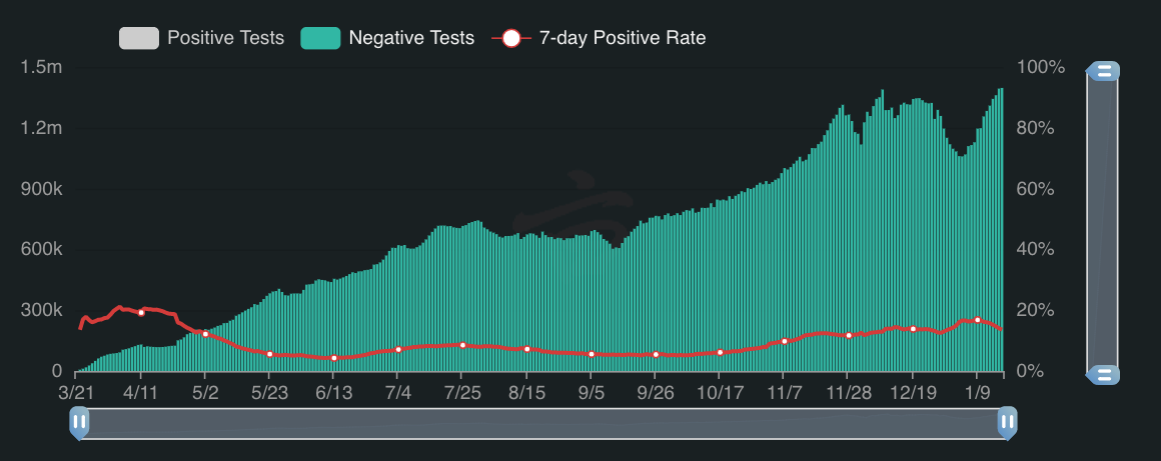}}
\subfigure[After reconfigure]{
\label{Fig.sub.2}
\includegraphics[width=0.35\textwidth, height=2cm]{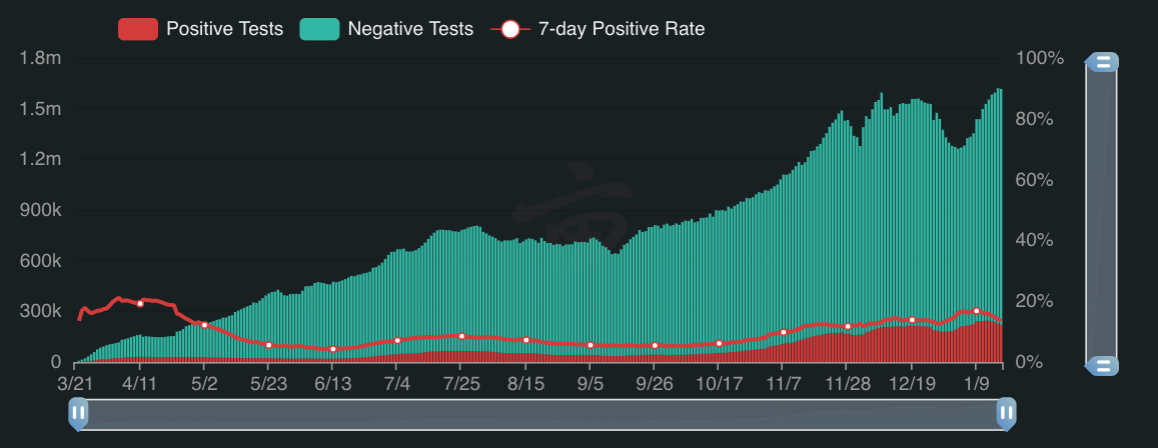}}
\caption{Reconfigure interaction}
\vspace{-4mm}
\label{Fig.reconfigure}
\end{figure}

According to Table~\ref{tab: data visualization and interaction}, we found that map, histogram, and line chart are the three most common visualization approaches,  applied to about 4/5 (80\%) dashboards that we collect. 
We also discovered that the most common interaction techniques are ``\textit{reconfigure}" and ``\textit{connect}". 
For example, in Fig~\ref{Fig.sub.1} and Fig~\ref{Fig.sub.2}, when the user clicks the positive tests button, both data and baseline have been changed.

In the second column,  the WHO dashboard and the JHU dashboard implement bubble maps to visualize data.
The bubble map applies all interaction techniques except ``\textit{select}" and ``\textit{encode}". 
Tableau data is also widely applied in those dashboards with the same proportion (80\%). 
3/4 of tableau data applies the techniques ``\textit{explore}" and ``\textit{connect}", and no table uses ``\textit{encode}" or ``\textit{filter}" for interaction.
Besides, three out of four tables provide the download data function.
The histogram is used in 80\% of dashboards we observed, and all of them use the ``\textit{reconfigure}" and ``\textit{connect}" function, but they have not used the ``\textit{select}" and ``\textit{encode}" interaction techniques. 
Donut chart and treemap are two visualizations that fewer dashboards used (2/5 donut chart, 1/5 treemap). Though they are rarely implemented, they can help users easily identify the partitions, and compare them with the whole objects.
Both of them use ``\textit{explore}" and ``\textit{reconfigure}" techniques to interact with users, but the donut chart also uses ``\textit{filter}" and ``\textit{connect}" techniques. 
All of those dashboards implement line charts. Different from static line charts, 80\% of linear charts used ``\textit{explore}", ``\textit{reconfigure}" and ``\textit{connect}" interaction approaches, which allows users to select the information that they are interested in and ignore the rest. 
Compare with line charts, fewer dashboards (40\%) develop bar charts with ``\textit{explore}", ``\textit{reconfigure}", ``\textit{filter}", and ``\textit{connect}" to interact, which may because line chart can better demonstrate the overall trend and is often used to represent changes in a time series. 

Additionally, we also noticed that the ``\textit{encode}" interaction technique is not applied in all of the eight visualizations we find, which may be due to the fact that the COVID-19 information is straightforward which doesn't require changing a type of representation for better understanding. 
Whereas the technique ``\textit{explores}", ``\textit{reconfigure}", and ``\textit{connect}" are the most common interaction techniques, which have been used in all of the visualizations.

\begin{center}
 \begin{tcolorbox}[colback=black!5!white,colframe=black!75!black,bottom=-0.05pt,top=-0.05pt]
According to the above analysis, we find that treemap, tabular data, and donut charts are used for one-dimensional data; line charts, bar charts, and histograms always deal with two-dimensional data; and bubble charts and maps can be adapted for multi-dimensional data types. 
We also determine that the interaction techniques ``\textit{explore}", ``\textit{reconfigure}", and ``\textit{connect}" are the most commonly used.
\end{tcolorbox}
\end{center}

\section{RQ4: Users' Comments on Current Dashboards}

In RQ4, we analyze comments from dashboards users. Based on the comments, we further summarize pros and cons of each dashboard, which shall help developers improve the dashboards and provide a better service.

\begin{table*}[]
\centering
\resizebox{0.8\textwidth}{!}{%
\begin{tabular}{l|c|ccccc}
\hline\hline
 &
  \multicolumn{1}{l}{} &
  \multicolumn{5}{c}{Keywords of Dashboard Comments} \\
  \hline
 &
  \multicolumn{1}{l}{} &
  WHO &
  JHU &
  1P3A &
  COVIDAu &
  DXY \\
  \hline
 &
  Influences &
  reminder, trustworthy &
  premier, authentic &
  best &
  excellent &
  \cellcolor[HTML]{FFFFFF}professional, Chinese \\
 &
  Usability &
  easy-to-use, brilliant &
  easy-to-use &
  - &
  easy-to-use &
  useful  \\
 &
  Data &
  \begin{tabular}[c]{@{}c@{}}comprehensive, real-time, \\ regional-data, accurate\end{tabular} &
  \begin{tabular}[c]{@{}c@{}}trend, aggregated,\\  latest, accurate\end{tabular} &
  \begin{tabular}[c]{@{}c@{}}easy-to-understand, \\country-level, accurate \end{tabular}&
  easy-to-understand &
  \begin{tabular}[c]{@{}c@{}}town-level,\\ prudent\end{tabular} \\
 &
  Visualization &
  good &
  helpful, interface &
  - &
  impressive, clear&
  - \\
\multirow{-5}{*}{Positive} &
  Others &
  SARS, mobile-friendly &
  easy-to-interact &
  mobile-friendly &
  mobile-friendly &
  aggregated \\
  \hline
 &
  Influences &
  - &
  - &
  - &
  - &
  nongovernmental \\
 &
  Usability &
  useless &
  dashboard-crashes &
  - &
  - &
  - \\
 &
  Data &
  \begin{tabular}[c]{@{}c@{}}lag, demographic, \\ racial, vaccine-tracking\end{tabular} &
  \begin{tabular}[c]{@{}c@{}}errors, lag, misleading, \\ double-counting\end{tabular} &
  lag, error &
  \begin{tabular}[c]{@{}c@{}} case-number,\\  inaccurate, error\end{tabular} &
  - \\
 &
  Visualization &
  \begin{tabular}[c]{@{}c@{}}mistake, marker, \\  hard-to-understand\end{tabular} &
  \begin{tabular}[c]{@{}c@{}}complex, distortion, \\ hard-to-understand\end{tabular} &
  hard-to-understand &
  \begin{tabular}[c]{@{}c@{}} geo-point, \\ hard-to-understand, map-design\end{tabular} &
  - \\
\multirow{-5}{*}{Negative} &
  Others &
  dashboard-management &
  content-modification &
  - &
  - &
  information removed\\
  \hline\hline
\end{tabular}%
}
\caption{Topic words from comments on Twitter}
\vspace{-6mm}
\label{tab:UserComments}
\end{table*}

\subsection{Approach}

As the developers of the COVIDAu dashboard, the authors have collected feedback about the dashboard from users to better understand users' needs and to try to modify the dashboard and provide the information they need. 
We set up a Google Form\footnote{\url{https://bit.ly/3uFPC4e}} for COVIDAu dashboard users to express their feelings about the dashboard and leave us some comments and suggestions. For the rest of the dashboards, we crawled related tweets from Twitter between 22${^{nd}}$ March to 10${^{th}}$ June 2020. 
We applied Twitter search API\cite{PythonTw97:online} to access Twitter data. 
We used the command line application to fetch the tweets which contain the dashboard names, e.g. \textit{JHU dashboard}, \textit{DXY dashboard}, etc, and set the language to English.
Similar to the approach in Section~\ref{sec:open-coding}, we then performed open coding in the following way. Two authors independently coded the comments for each dashboard to label the comments as \textit{``Positive"} or \textit{``Negative"} and we stored all the information in a table\footnote{Data release at URL: https://cutt.ly/Kjuzb5J}.
After the initial coding, the two authors discussed any discrepancies and comments categories until a consensus had been reached. 

\subsection{Results}

\textbf{Comments on COVIDAu Dashboards}

In total, to date we have received 347 feedback messages from our dashboard users. 
Nearly 35\% of them  express their appreciation of the dashboard. However, 32\% of them  requested new information related to the pandemic and data visualizations. 
% We discuss the users' requests and compare them with the findings we summarized in RQ1 and RQ3 in this section.

The requests we received are categorized into the topics we summarized in RQ1. 
For example, ``\textit{Could you do a breakdown of local transmission?}" is asking for the transmission information of COVID-19 (Topic 1 Table~\ref{tab:twitterTopics}), ``\textit{Need some context in terms of why flights are posted.}" is related to international travel (Topic 14 Table~\ref{tab:twitterTopics}), and ``\textit{Can you please update the information regarding current regulations}" is about government policies (Topic 16 Table~\ref{tab:twitterTopics}). 
We found all of the requests are associated with COVID-19 information (Theme 1) and impact on society (Theme 3). 
% This suggests that the requirements we extracted from tweets are matching to  users' real information needs.

However, there are no requests that are related to the protection approaches (Theme 2) and impact on business (Theme 4). 
This may be because, for theme 2, the COVIDAu dashboard has covered most of these topics and the approaches haven't changed since the start of the pandemic. 
Therefore, our users didn't request such information. 
For the impact on small business, as we will discuss in Section~\ref{sec:discuss_information_gap}, no current COVID-19 dashboards that we analyzed provide any such information. 
This may indicate that though the public is seeking this type of information online, they may not realize that content of impact on small business can be presented on dashboards. Hence, we haven't received feedback ask for information related to theme 4.
% To justify the theory, we plan to add such information on the dashboard and see how the users react to it in the future works.

There are also requests that are related to data visualizations used. 
For example, \textit{``I would love a \"new cases\" bar graph on the state pages"} and \textit{``Really good job ... The only thing I would like to see added is a state filter on the new cases per day bar graph"}, the two requests are asking for a bar chart visualization on the new cases data. 
Moreover, requests like \textit{``You should not have map points to people's homes. Not everyone is going to click and see that it is not a location of an actual confirmed case."} is suggesting not to use markers to visualize the confirmed case. 
Instead, all dashboards we selected use layers or bubble charts on the map to visualize such data, which helps users to avoid misunderstandings.
% We plan to conduct more research to justify the relation between users' comments and their dashboard-based information needs in future work.

\textbf{Comments from Twitter}

According to Table.~\ref{tab:UserComments}, the comments mainly concentrate on four aspects: influence, usability, data, and visualization.
More than half of the people(63\%) gave positive comments to the dashboards.
These comments mainly elaborated on the important role that dashboards play in global epidemic control.
Some of them talked about how the interface is user-friendly, the data is comprehensive, accurate, and real-time updated, and the visualization of graphs helps them understand the pandemic.
For example, the comments \textit{``This \#JohnsHopkins  dashboard is proving to be one of the more reliable sources of information about the spread of coronavirus \#covid19"} affirmed the credibility of the JHU dashboard and the comments \textit{``WHO had a central dashboard with such figures which were up-to-date. There is a lot of value in it."} discussed the data on the WHO dashboard is comprehensive.

There are also negative comments about dashboards as well.
According to Table.~\ref{tab:UserComments}, those tweets mainly focus on criticizing data accuracy, visualization, etc.
14\% of the comments think that the dashboard data is inaccurate, lagging, and lacks certain types of data they need (e.g., \textit{``WHO's dashboard on its website which tends to lag countries' individual tallies showed that Monday."} and \textit{``Quick question :  Why not update the coronavirus dashboard for Yemen?"}).
% For inaccurate data and untimely updates, this may be caused by the different times when the information is released in different countries/regions.
The inaccurate data and untimely updates could be related to different data sources and time zones.
For the lack of information that users need, this may be because the developer does not realize that people need this type of information, or only a small portion of people need this type of information.
About 6\% of users also raised negative comments about visualization, such as errors in the map.
For example, \textit{``If as stated Baw Baw region has confirmed cases, why is the geo point very definitely at my property in Paynters Road Hill End? It should be at headquarters in Warragul.... Please adjust if you want to be accurate."}. 
This may be due to the developers not being familiar with the map information or local customs and not being aware of the problems. Users seem to have diverse positive opinions on different dashboards. However, for negative comments, users tend to raise similar problems. 
For example, users complain that the visualizations are hard to understand on four out of the five dashboards. 
Besides, comments about data errors are also common among different dashboards.
For instance, users posted  \textit{``...There's some serious problem with data gathering. ... How does the WHO have better data than the states?"} and  \textit{``Number of COVID-19 deaths in the U.S. as of yesterday according to the latest Johns Hopkins University (JHU) data dashboard (with likely under-reporting of death numbers)."}.

\begin{center}
 \begin{tcolorbox}[colback=black!5!white,colframe=black!75!black,bottom=-0.05pt,top=-0.05pt]
While most users have positive feedback to the current dashboards, they also raised their concerns online as well. It's common to see users report errors or bugs, request for certain types of data, and give suggestions on data visualization. 
For different dashboards, the positive comments often contain unique information, whereas the negative comments are similar under the categories. 
\end{tcolorbox}
\end{center}

\section{Discussion}

In this section, we compare people's information needs with dashboard supply and discuss the gap between the two.
In addition, we also analyze the visualization and interactions on dashboards and summarise lessons learnt from the study.

\subsection{The Gap between People's Needs and Information Supply}

\begin{figure}[h]
\centering
\includegraphics[width=0.43\textwidth]{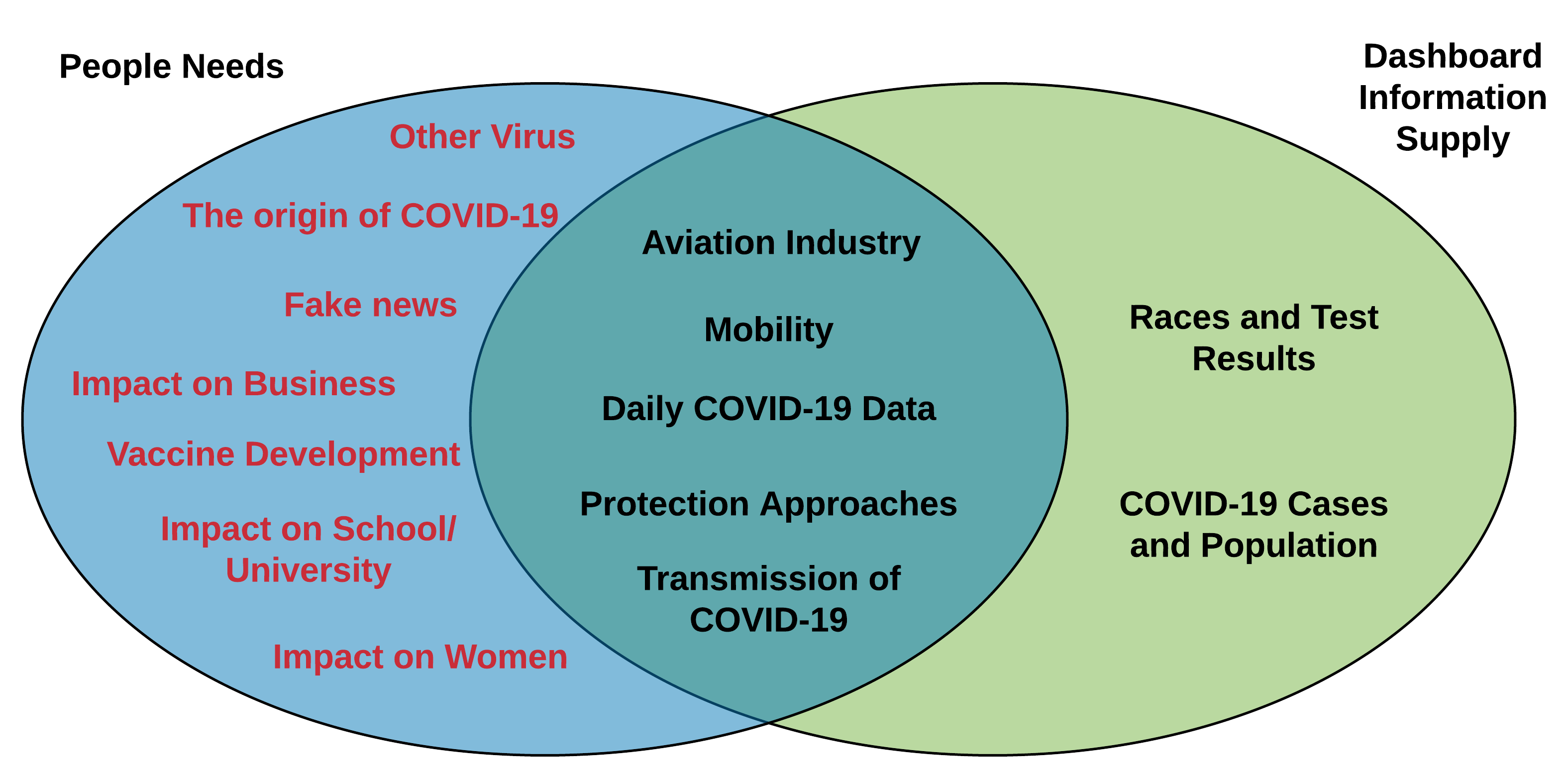} 
\vspace{-2mm}
\caption{Information gap between people's needs and dashboard supply} 
\vspace{-3mm}
\label{fig:gap} 
\end{figure}

Our research finds that people are more concerned about COVID-19 information, protection approaches, society issues, and business impact. 
However, judging from the existing dashboards we analysed in RQ2, there is still a gap between the information on dashboards and people's needs, shown in Fig~\ref{fig:gap}. 
Most of the dashboards we investigated include the impact of COVID-19 on people, such as confirm cases, death. 
This part also contains protective measures and methods, such as a correct hand-washing video. 

\label{sec:discuss_information_gap}

Nevertheless, no dashboards cover the impact on the business dimension like a series of government policies for SMEs (small and medium enterprises) and some topics we present in Table~\ref{tab:twitterTopics}(ID: 2,3,4,7,17,18,19). 
We summarise three reasons why the above information is absent. 
One reason is that data is challenging to visualize, for example, data with high dimensions like economic market data.
Economic market fluctuations are affected by many factors, such as government fiscal and monetary policy and natural disasters/extreme weather fluctuations~\cite{hall_2021}. 

Another reason is that such data can be hard to collect. 
For the above missing information, developers can collect local labour market information during the epidemic for visualization and then reflect the impact of the epidemic on the economic market. 
Similar to the 1Point3Acres website, they use the number of people who declared unemployment insurance during COVID-19 to show the impact of the epidemic on the unemployment rate, which is also a reflection of the COVID-19's impact on society.

The last reason is  that perhaps the developers didn't realize that people need this information.
Most dashboards we analysed focus on general COVID-19 information, which is the majority of users' essential information needs.
Some information needs are for a specific group. For example, business-related information is more critical for business owners so that the developers may not realize this type of information needs.

\begin{table}[!htbp]
\centering
\begin{tabular}{c|c}
\hline
Visualisation (actions) & Portion                          \\
\hline
Map (zoom, click, span)  & 71.70\%        \\
choropleth map (choose state)  & 10.50\% \\
Linear chart (select countries) & 7\%  \\
Tabular data search (search)                       & 1\% \\
\hline
\end{tabular}
\caption{User actions in COVIDAu}
\vspace{-5mm}
\label{tab:User_actions}
\end{table}

\begin{figure}
\centering 
\subfigure[Select countries on COVIDAu]{
\label{fig:selectCountries}
\includegraphics[width=0.23\textwidth]{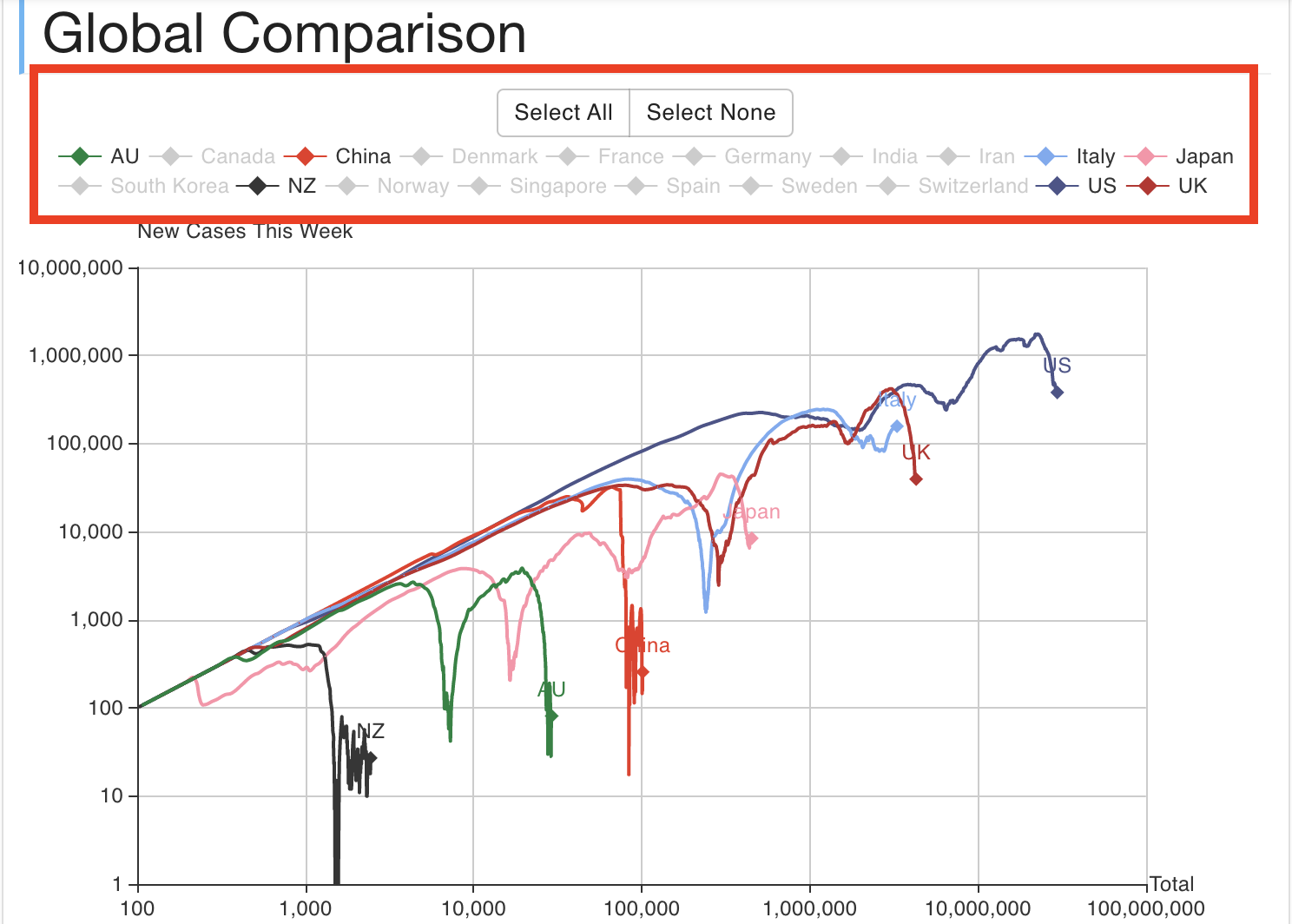}}
\subfigure[Flights on tabular data]{
\label{fig:tabular}
\includegraphics[width=0.23\textwidth]{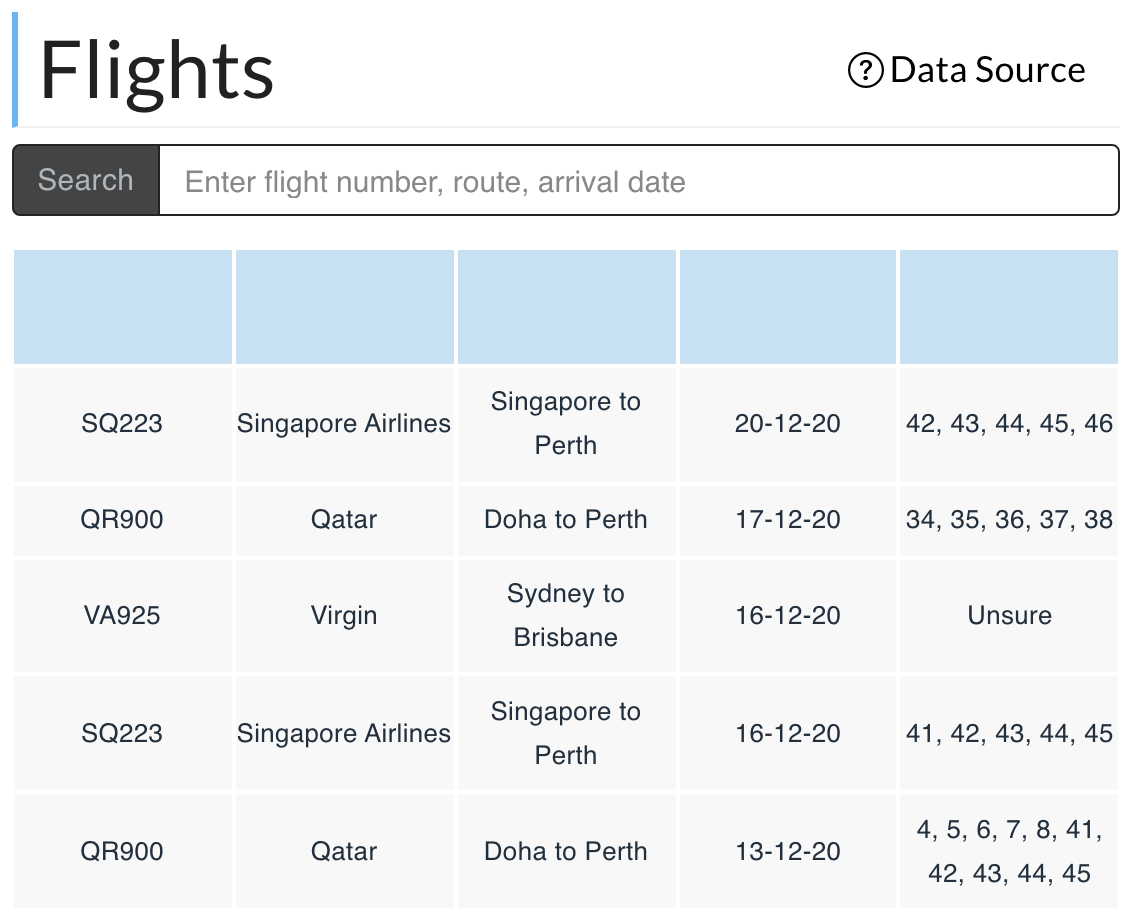}}
\vspace{-4mm}
\caption{Examples for discussion}
\vspace{-6mm}
\label{fig:Examples}
\end{figure}

\subsection{Data Visualization and Interaction}

In terms of data visualization types used, we find that dynamic visualizations are widely used compared with static charts when observing the changing data, as they can make it easier for users to understand~\cite{suyatna2017role}. 
Our research team collected data about user activity on our dashboard to analyze which visualization is frequently used.
% After analyzing user actions from the COVIDAu website in Table.~\ref{tab:User_actions}, we found that map visualization is more popular with users because 71.7\% of users used this visualization 
After analyzing users behavior on the COVIDAu website, we found that the map visualization is the most popular one as 71.7\% of users used this visualization, seen in Table.~\ref{tab:User_actions}.
Users do not often interact with tabular data search, and this may be because the map information will be more obscure and require users to further explore through some interactive operations. 
However, tabular data is more intuitive, so relatively few user interaction operations are required. 

Visualizations can reduce the cognitive cost of inspecting, but sometimes it depends on the user's cognitive styles and task difficulties.
Luo's research~\cite{LUO2019113061} finds that users' cognitive styles have significant effects on how they choose the visualization format.
For example, a table format is always chosen by users with verbalizer cognitive style.
A complex data visualisation type can increase the difficulty of the user's understanding.
For example, consider the select countries operation in the linear chart Fig~\ref{fig:selectCountries}, 7\% users use this interaction because many users don't know that the country on the picture can be selected and compared with the drop down list.
This design is more complicated and challenging to use. 
Similarly, for tabular data in Fig~\ref{fig:tabular}, users prefer to see the data they want intuitively. 
That's why there are only 1\% of users used the search function.

Beside the COVID-19 dashboard, we also compare the findings in data visualization and interaction with other crisis event dashboards.
On the NASA Disasters Program Response Locations dashboard~\cite{NASADisa46:online}, it is clear that the line charts and bar charts contain two-dimensional data (time and event), and developers use ``\textit{connect}" interaction approach to help users with a better understanding. 
The map visualization on this dashboard uses ``\textit{explore}" and ``\textit{connect}" approaches, and displays multi-dimensional data (time, location, and event) to users.
Another dashboard for Flow and Drought Management in the Delaware River Basin~\cite{HydroRep72:online} also shows two-dimensional data on both line charts (time and flow) and bar charts (location and usable storage).

\subsection{Lessons Learnt on Data Visualization}

We summarize three lessons learnt from mistakes made by the dashboards on data visualizations. The lessons may be helpful for developers who plan to develop a dashboard, and bring a better web service to the audience.

\textbf{Visualize Sensitive Data with Caution}
During the development of the COVIDAu dashboard, we used small red markers to represent confirmed cases in certain suburbs. The markers were set in the center of the suburb maps by default, which caused problems that they may be set on the top of someone's property or in a desert. These were so confusing to the public that we received complaints and questions constantly. We then replaced it with different shades of blue to represent confirmed cases in a certain area to avoid the ambiguous and misunderstanding. Every number behind the map represents human lives, when visualizing this type of data(sensitive), developers or website designers should be more careful about the types of chart/map to use to provide many accurate and useful data to the public.

\textbf{Annotate Complex Visualizations}
 From the users' perspective, some of the data visualizations are not easy to understand. For example, as shown in Fig~\ref{fig:selectCountries}, the concept of this chart was borrowed from this site~\cite{web:trajectorychart}, and similar charts have been found in two of the dashboards. The chart uses the logarithmic scale of total confirmed cases and the newly confirmed cases per week as X and Y axis to better demonstrate the trend of the pandemic. Since dashboards users are from different age groups and backgrounds, it's hard for some of them to understand the meaning behind it. The authors also received a number of requirements from users asking for instructions on how to read the chart when developing the COVIDAu. Therefore, proper annotations for the visualizations and domain-specific terms are required to help users process the information on dashboards.

\textbf{Use Professional and Consistent Terms}
We noticed that different dashboards may have different terms for the same concept that could confuse users. For example, the COVIDAu dashboard used to have \textit{Existing Case} to represent the cases that have been confirmed but haven't recovered or died, whereas other dashboards were using \textit{Active Case}. The inconsistency in wording has triggered users to question the professionalism of the dashboard, and concern about the data accuracy. Similar situations were also identified in other dashboards. To help users understand information on dashboards, software engineers need to use professional terms instead of synonym words in data visualization.

In this work, we confirm the gap between people's information needs and dashboard information supply. 
We also find the main types and patterns of data visualization and interaction on dashboards and summarized lessons learnt from the study.
However, there are several issues that require further improvement, such as how to improve the accuracy in tweet question detection. 
%We leave these questions as our future work.

\section{Threats to Validity}

We summarized three validity threats that can be bias of our empirical study.

\textbf{Construct validity:} is related to the metrics and measurement procedures we used to code the dashboards and extract information needs.
Most constructs that we apply are defined and used in previous research, i.e.,
the question detection~\cite{GitHubka30:online} and
LDA topic modeling~\cite{tong2016text}.
% In order to reduce the error of topic modeling, we adopt a question detection process by filtering out the tweets that contain the questions, and then we start topic modeling.

\textbf{Internal validity:} is related to evolution of dashboard development, tweets trends changing in different stages of the COVID-19 pandemic, and the human bias when collecting information supplies from dashboards.  
In an online environment, dashboards are consistently updated and improved. 
To mitigate the dashboard evolution threat, we regularly update our information supply sheet and use the latest version for experiments.
Regarding the changing trends of information needs on social media, we will keep collecting data and prepare for future research.
All the information supplies we identify are collect manually, which is can be a bias to the result. 
% We apply an open-coding approach to mitigate this threat, and two groups of our researchers work on the problem .

%External validity, related to the selection of the tweets and coding dashboards. 
\textbf{External validity:} measures how much the results of our study can be generalized.
One threat to external validity is the limitation of the tweets and dashboards in our study. 
In our approach, we only select the English tweets and observe five dashboards.
To mitigate this threat, we can use auto-translate tools or machine learning models~\cite{chen2016learning} to include tweets from other languages and also add more popular dashboards from a broader areas.
Another threat is whether the findings can be generalized to other dashboards outside of our research.
To mitigate this threat,  we obtain dashboards from diverse sources that contain both local and global dashboards and originate from various countries.
% However, we make the complete information and visualization summary available online\footnote{Google sheets URL: https://cutt.ly/Kjuzb5J}.

\section{Related Works}

We briefly discuss some efforts that have been made by developers to deal with the COVID-19 pandemic, and some work related to dashboard development.

\subsection{Developers' Efforts in Battling COVID-19}

Pellert et al.~\cite{pellert2020dashboard} build a self-updating sentiment monitor by retrieving data from the news platform, Twitter, and a forum for young adults in Austria.
Chrysler et al.~\cite{chrysler2021mobile} designed a mobile application to track the geolocation of patients, which could help people avoid being in contact with the COVID-19 patients.
Davalbhakta et al.~\cite{davalbhakta2020systematic} performed a review of mobile apps related to COVID-19 and assessed their qualities.
% The researchers apply Linguistic Inquiry and Word Count dictionary for text analysis and classify emotions in social media.  
% This study identifies a series of events in Austria corresponding to anxiety emotions on social media, for example, the report of the first COVID-19 case and announced bans of large local events. 
Kabir and Madria~\cite{kabir2020coronavis} develop a real-time COVID-19 tweet analyzer, CoronaVis, which tracking, collecting, and analyzing tweets related to COVID-19.
% This web application contains topic trends, sentiment analysis, and user movement information on Twitter. 
% They use an LDA algorithm and pyLDAvis package to produce an interactive visualization of each topic's top-30 most relevant words. 
% Nguyen et al.~\cite{nguyen2020wnut} organized a competition for participants to submit high-performance models to perform classification and sentiment analysis on COVID-19 tweets to help downstream applications select the informative Tweets.
% Haggag et al.~\cite{COVID19privacy} compare the privacy issues on COVID-19 apps versus social media apps. 
%In terms of virus detection, Andreu-Perez et al.~\cite{Andreu-Perez2021} propose the DeepCough3D that can offer private screening COVID-19 test services by analyzing respiratory sounds with deep learning methods.
In terms of topics related to COVID-19 in Twitter, Zheng et al.~\cite{uncovering_topics} discussed popular topics among Twitter users from 11 March to 25 March 2020 and Asgari-Chenaghlu et al.~\cite{asgaricovid} applied natural language processing to detect the trending COVID-19 related topics in Twitter. 
Chang et al.~\cite{chang2021people} and Zhang et al.~\cite{zhang2021understanding} explored people's concerns on Twitter during the COVID-19 pandemic, but Zhang et al.'s study also performed sentiments and disparities analysis. 

Similar to Pellert et al.'s research~\cite{pellert2020dashboard}, our research also checks word distributions from the database and extracts the requirements with high accuracy.
% Specifically, we also generate keywords in each topic, which is easier to read and perform cleaning tasks.
So far, no work seems to have been done on the gap between COVID-19 dashboards information supply and people's information needs. 
Our work investigates the information and visualization on dashboards and compares it with people's questions related to COVID-19 information seeking on Twitter.
% However, future work must better address data volume.

\subsection{Dashboard Development}

Developers have invented different types of information dashboards in various domains response to problems such as slow information dissemination or misinformation.
The information dashboard, as an emerging tool, has been explored to update real-time data and visualize data.

From 2003, dashboards like~\cite{ArcNewsS67:online, WHOEbola72:online} have been developed to serve the public needs for virus like SARS, Ebola, and Zika.
% used dashboard tracking SARS in China with GIS~\cite{ArcNewsS67:online}.
% They developed a dashboard that could analyze information and disseminate information to residents, and benefit the world.
%The system is based on ArcIMS and has two interfaces in both Chinese and English.
%This website provides daily up-to-date information about the distribution of diseases in Hong Kong, China, and the world.
% Also, it is presented in the form of an online map and formed a prototype of the dashboard.
% WHO developed a dashboard in 2015~\cite{WHOEbola72:online},  to report the Ebola situation reports in Guinea, Liberia, and Sierra Leone. 
% The reports contain geographical distribution of new and total confirmed cases, death number, confirmed Ebola cases, and Ebola treatment centers' location.
% In 2016, NASA presented a dashboard that helps forecast Zika risk~\cite{NASAHelp96:online}.
% The agency works on predicting the potential spread of the Zika virus in the United States.
%Their Zika risk map-enabled governments and health organizations better prepare for disease outbreaks caused by the spread of the virus.
% This dashboard uses a bubble map in visualization, and it provides a detailed explanation for elements on the bubble map
% Nowadays, in response to the ongoing public health event, the Centre for Systems Science and Engineering at Johns Hopkins University~\cite{dong2020interactive} develop a real-time dashboard and share it publicly. 
% The dashboard reports cases in several countries but at different levels, such as the province level in China and Australia's city level. 
Nowadays, in response to the ongoing COVID-19 pandemic, Boulos and Geraghty~\cite{boulos2020geographical} performed research on map-based dashboards such as the dashboard developed by JHU to identify types of data that is presented on the map and discuss the limitation of the JHU dashboard.
Arneson et al.~\cite{arneson2020covidcounties} developed a real-time COVID-19 tracker at the US level for people without a data science background. 
They created several novel visualizations, which include local cases doubling times and estimated ICU requirements by county.
Yoo and Kronenfeld~\cite{il2019evaluation} carried out research to figure out the best visualization approaches for the epidemiological data during the COVID-19 pandemic. They raised questions about data variables, visual variables, map interaction, and dashboard response speed.

%JHU CSSE's dashboard as a case in point, their work identified the data presented on the interactive map, such as confirmed infections and fatalities.
% % Moreover, the researchers also demonstrate the limitation of this dashboard, which could not archive previous data.
% Another work from Wimba et al.~\cite{wimba2020dashboard} created a dashboard to track the COVID-19 outbreak in the Democratic Republic of Congo. 
% In this dashboard, the percentage of people with masks and confirmed cases number have been recorded. 
%It also proved that a dashboard could be created with limited human resources and technical resources. 
%This result will encourage researchers with a limited budget to create a dashboard like this to fighting the COVID-19 pandemic.
% Sharma et al.~\cite{sharma2020coronavirus} designed a dashboard to track and help the public understand misinformation on social media. 
% This dashboard provides the identified misleading and low-quality information spread on social media. 
% Furthermore, this kind of information has been divided into four categories, which are unreliable, conspiracy, clickbait, and political/biased. 
% Researchers also publish those low-credibility tweets on their dashboard and analyze the spreading across countries. 

There is an increasing number of dashboards developed by various countries to help manage COVID-19 pandemic issues.
Wissel et al.~\cite{wissel2020interactive} aggregate data and public an interactive dashboard in the US. 
The COVID-19 tracker~\cite{COVIDtracker} and Daniel Conlon's dashboard~\cite{Coronavi62:online} covers global COVID-19 information. 
% Wimba et al.~\cite{wimba2020dashboard} created a dashboard to track the COVID-19 outbreak in the Democratic Republic of Congo. 
%European Centre for Disease Prevention and Control also release a dashboard~\cite{Homepage79:online} to track real-time COVID-19 data in Europe. 
%Besides, the University of Melbourne developed a dashboard for COVID-19 10-day forest~\cite{COVID19MEL}.
Dashboards can provide front-end visualizations and data management services, the works below also explore these two services.
Chen et al.\cite{5928315} talked about visualization design on the interface to perform a better understanding of the service performance. 
They provided the recommendations visualizations and the steps of drawing the visualization.
Zheng et al.\cite{6357180} conducted an evaluation on real-world web service, collected and released web service QoS data sets for further research.
Another work\cite{upadhyaya2015quality} explored perceived quality from the user’s perspective for service selection and composition. They named this quality as quality of experience (QoE).
The above works focus on building a dashboard or analyzing some specific elements on front-end and data management services.
Our work covered several dashboards, and we applied open coding strategies to collect data on each dashboard.
We also analyzed the data provided, visualization approaches and interaction techniques used on these dashboards to summarise the key usage patterns among them.

\section{Conclusion and Future Work}

The information dashboard has become an essential tool to make sense of ever-changing data in this data-driven world. 
Thus, ensuring that the information supplies actually serve people's needs has become a priority for dashboard design and development. 
In this research, we explore the information needs of COVID-19 dashboard users and conducted a study of five COVID-19 information dashboards. 
We conducted tweet analysis and collected information from dashboards using an open coding approach. 
We identified four key dimensions of information on COVID-19 dashboards: location, human, time, and others. 
We further explained sub-dimension information that should be considered when developing a crisis management and information dashboard. 
We illustrated the variety of visualization types and interaction mechanisms that these dashboards support. 
We determined the current gap between people's information needs and information supplied on COVID-19 dashboards based on our data analysis.  
To make visualizations more accessible for future crisis dashboard development, developers need to use appropriate visual designs to present different information types and need to support suitable interaction techniques to navigate using different visualizations. We also summarized lessons that shall prevent developers from making the same mistakes in the future.

As our future work, we plan to validate and extend the findings by studying a more extensive set of dashboards with more diverse visualizations and interaction dimensions. 
Furthermore, we plan to explore the approaches of mining information needs with higher accuracy. 
We also want to support other types of information analysis such as the misinformation topic detection during the pandemic.

\section*{Acknowledgements}

Grundy is supported by ARC Laureate Fellowship FL190100035.

\bibliographystyle{IEEEtran}
\bibliography{reference}

\vskip -4\baselineskip plus -2fil 
\begin{IEEEbiography}[{\includegraphics[width=0.85in,height=0.98in,clip,keepaspectratio]{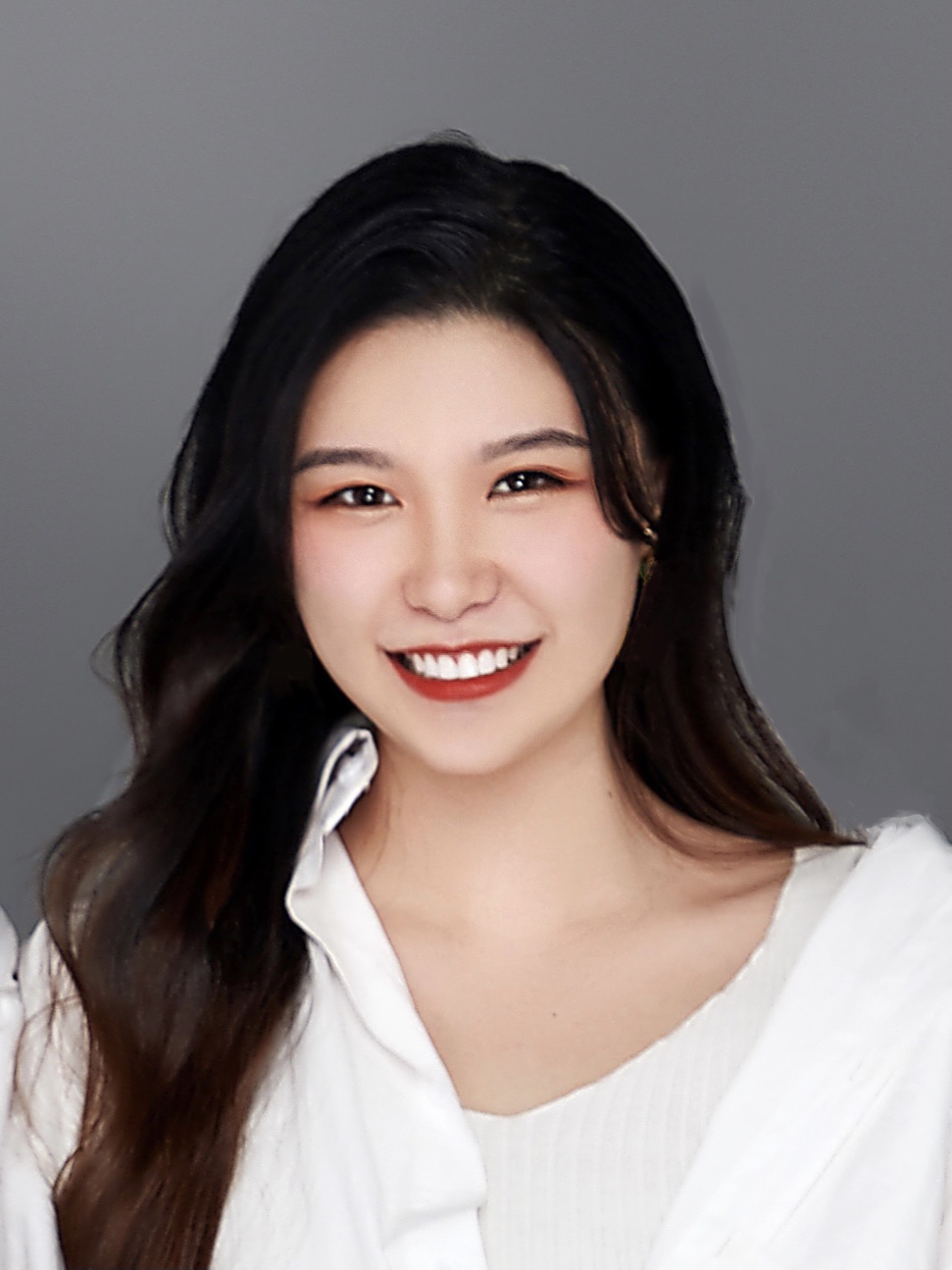}}]{Xinyan Li}
graduated from Southwest University, China in 2017. She is a master student from the Faculty of Information Technology, Monash University, Australia supervised by Chunyang Chen. Her research focuses on data analysis such as data mining and machine learning.
\end{IEEEbiography}
\vskip -5\baselineskip plus -2fil 
\begin{IEEEbiography}[{\includegraphics[width=0.85in,height=0.98in,clip]{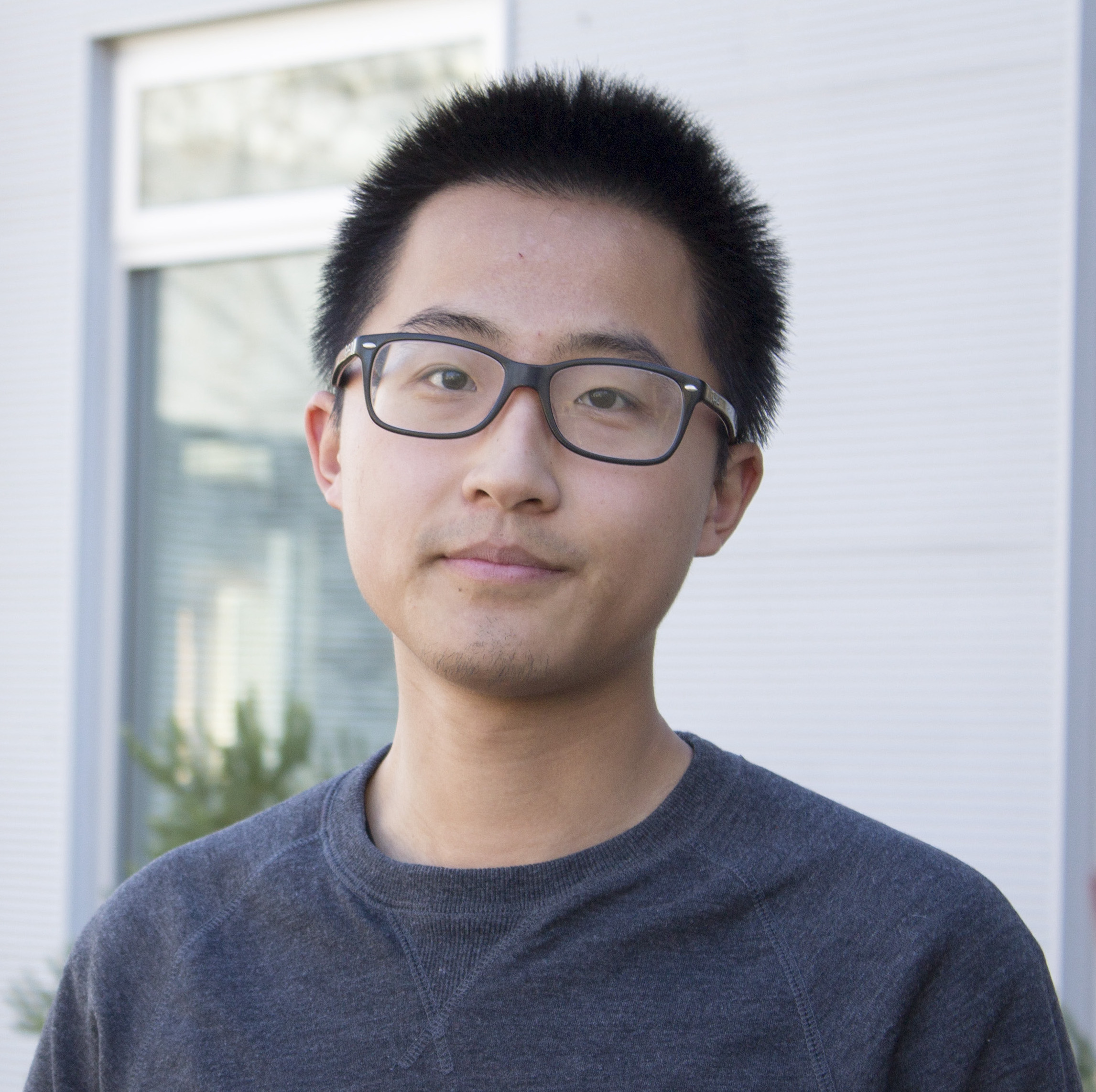}}]{Han Wang}
received the B.S. degree with honours in Software Engineer from Australian National University, Australia, in
2017. He is currently a PhD student from Faculty of Information Technology, Monash University, Australia supervised by Chunyang Chen. He focuses on applying techniques such as deep learning, NLP, etc., to help developers and end users.
\end{IEEEbiography}
\vskip -5\baselineskip plus -2fil 
\begin{IEEEbiography}[{\includegraphics[width=0.85in,height=0.98in,clip,keepaspectratio]{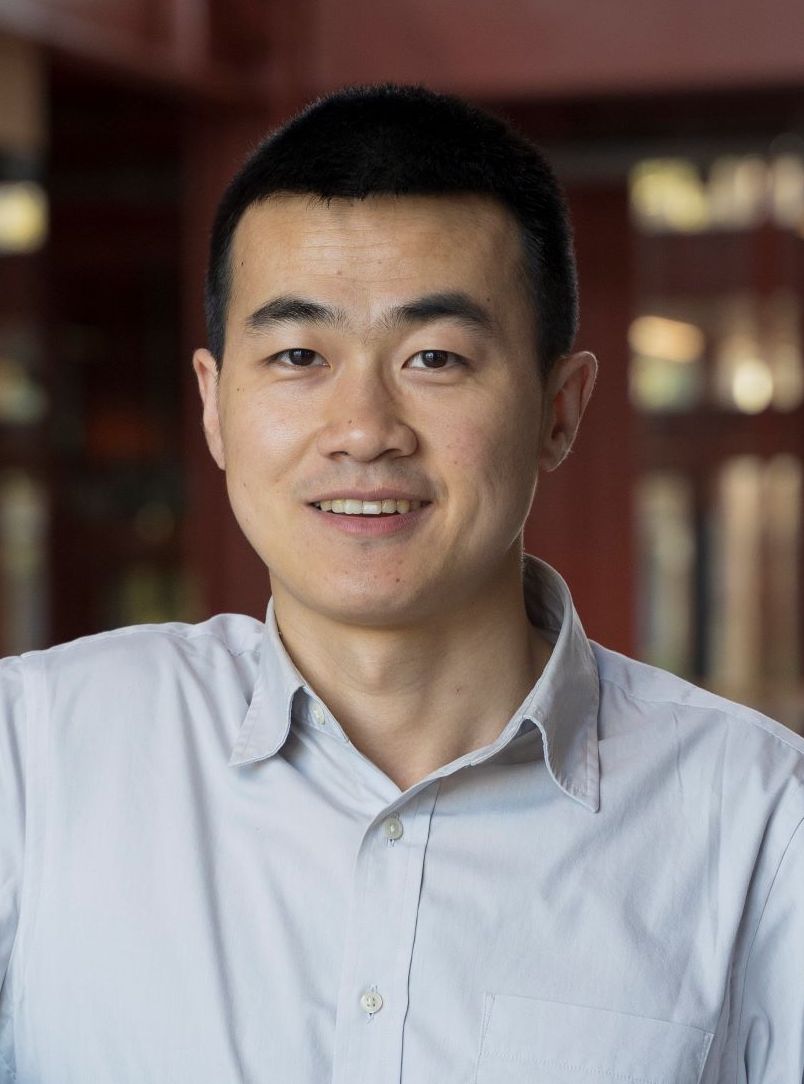}}]{Chunyang Chen}
is a lecturer (Assistant Professor) in Faculty of Information Technology,
Monash University, Australia. His research focuses on software engineering, deep learning
and human-computer interaction. He has published over 50 papers in referred journals or conferences.
His research works are well recognized by best paper awards including 3 ACM SIGSOFT Distinguished Paper Awards in ICSE'21, ICSE'20 and ASE'18. 
\end{IEEEbiography}

\vskip -5\baselineskip plus -2fil 
\begin{IEEEbiography}
[{\includegraphics[width=0.85in,height=0.98in,clip,keepaspectratio]{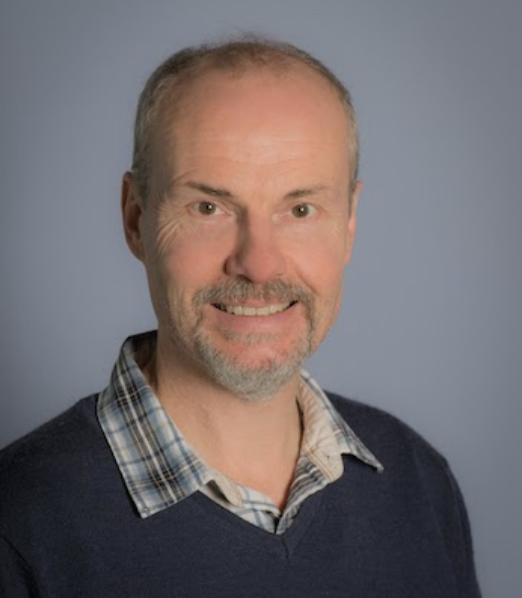}}]{John Grundy}
is Australian Laureate Fellow and Professor of Software Engineering at Monash University, Australia. He has published widely in automated software engineering, domain specific visual languages, model-driven engineering, software architecture, and empirical software engineering, among many other areas. He is Fellow of Automated Software Engineering and Fellow of Engineers Australia.
\end{IEEEbiography}

\end{document}